\documentclass[12pt]{article}
\usepackage{amsmath} 
\usepackage{amssymb} 
\usepackage{slashed} 
\usepackage{color}
\usepackage{graphicx}
\usepackage{caption}
\usepackage{subcaption}

\usepackage{times}

\title{Effective-field theory analysis of the $\tau^{-}\rightarrow (K \pi)^{-}\nu_{\tau}$ decays} 

\author
{Javier Rend\'on,$^{1}$ Pablo Roig$^{1}$, Genaro Toledo,$^{2}$\\
\\
\normalsize{$^{1}$Departamento de F\'isica, Centro de Investigaci\'on y de Estudios Avanzados}\\
\normalsize{del IPN, Apdo. Postal 14-740, 07000 Ciudad de M\'exico, M\'exico.}\\
\normalsize{$^{2}$Instituto de F\'isica, Universidad Nacional Aut\'onoma}\\
\normalsize{de M\'exico, AP20-364, Ciudad de M\'exico 01000, M\'exico.}\\
}

\date{}

\usepackage{graphicx}
\begin{document}

\baselineskip24pt

\maketitle 

\begin{abstract}
\noindent We analyze the $\tau^-\to(K\pi)^-\nu_\tau$ decays within an effective field theory description of heavy new physics (NP) modifying the SM left-handed weak charged current and include refined SM input for the participant meson form factors exploiting chiral symmetry, dispersion relations and (lattice) data. We include the leading dimension six operators and work at linear order in the effective couplings. Within this setting we:\\
i) follow the derivation in Phys. Rev. Lett. 120 (2018) no.14, 141803 where it was proved unambiguously that it is impossible to understand the BaBar anomaly in the CP asymmetry measurement within this framework. We allow for reasonable variations of the hadronic input involved and study the associated uncertainty;\\
ii) first show that the anomalous bump present in the published Belle data for the $K_S\pi^-$ invariant mass distribution close to threshold cannot be due to heavy NP;\\
iii) first bind the heavy NP effective couplings using $\tau^-\to(K\pi)^-\nu_\tau$ decays and show that they are competitive with those found in hyperon semileptonic decays (but clearly not with those obtained in Kaon (semi)leptonic decays for NP scalar currents). We put forward that the comparison of the considered tau decays with (semi)leptonic kaon and hyperon decays provide with meaningful tests of lepton universality for (NP) tensor interactions.\\
We also compare the SM predictions with the possible deviations caused by NP in a couple of Dalitz plot distributions, in the forward-backward asymmetry and in the di-meson invariant mass distribution and discuss the most interesting measurements to be performed at Belle-II using these decays data.
\end{abstract}

\section{Introduction}\label{sec_Intro}
Even before its discovery \cite{Perl:1975bf} by the group lead by Martin Perl at SLAC, the tau lepton was identified as a very useful probe for precision electroweak studies and a clean QCD laboratory at low energies, thanks to the pioneering work of Yung-Tsu Tsai \cite{Tsai:1971vv}. The precise measurements achieved mainly thanks to the $Z$ and $B$ factories (see, for instance, Refs. \cite{Schael:2005am, Bevan:2014iga}) have promoted tau Physics to precision tests of the Standard Model (SM) implying non-trivial restrictions on its possible extensions \cite{Pich:2013lsa}.\\
In this article we focus on the $\tau^-\to (K \pi)^- \nu_\tau$ decays, that we study within the SM and considering the effects of heavy new physics (NP) on a number of phenomenologically interesting observables. A clear motivation for this is the BaBar anomalous measurement of the CP asymmetry in the $K_S\pi^-$ channel \cite{BABAR:2011aa}. The CP violation present in the SM \cite{Kobayashi:1973fv} is clearly insufficient to understand the baryon asymmetry of the universe \cite{Sakharov:1967dj, Cohen:1993nk, Riotto:1999yt} so that any hint of NP involving CP violation becomes a candidate for providing with a clue to understand the enormous matter-antimatter imbalance. With respect to this BaBar anomaly, however, the related Belle measurement \cite{Bischofberger:2011pw} of a binned CP asymmetry in the same decay channel analyzing the decay angular distributions is compatible with zero, as expected in the SM with a permille level precision. On the theoretical side, Ref.~\cite{Cirigliano-Crivellin} proved that heavy NP cannot explain this anomaly, as we discuss in detail when we list the main motivations for this work later in this introduction.\\
After the discovery \cite{Aad:2012tfa, Chatrchyan:2012xdj} and precise characterization \cite{Aad:2015zhl, Khachatryan:2016vau} of the Higgs boson at LHC, the electroweak sector of the SM does not show any hint of NP close to the top quark mass scale \cite{Baak:2012kk}. This confirms the usefulness of considering the SM as a low-energy effective field theory (EFT) of the underlying dynamics at some TeVs, the so-called SMEFT \cite{Buchmuller:1985jz, Grzadkowski:2010es} (when the spontaneous electroweak symmetry breaking is linearly realized). In this spirit, and for weak charged current decays involving light quarks, the low-energy limit of SMEFT is a Fermi-type theory with straightforward connection to the TeV physics if this is weakly-coupled \cite{Cirigliano_Jenkins_Gonzalez, Bhattacharya_Cirigliano, Cirigliano:2012ab, Cirigliano:2013xha, Chang:2014iba, Courtoy:2015haa, Gonzalez-Alonso:2016etj, Gonzalez-Alonso:2016sip, Alioli:2017ces, Gonzalez-Alonso:2017iyc, Gonzalez-Alonso:2018omy}. In this context, several recent works \cite{Garces:2017jpz, Miranda-Roig, Cirigliano:2018dyk} have put forward that semileptonic tau decays can be complementary to other low-energy semileptonic probes considered before: such as the corresponding pion and Kaon decays, nuclear beta or hyperon decays. In this paper we extend this latter effort to a complete phenomenological analysis of the $\tau^-\to (K\pi)^-\nu_\tau$ decays, which has not been done, within this framework, before.\\
Specifically, the main motivations for this study are:
\begin{itemize}
 \item Ref. \cite{Cirigliano-Crivellin} disproved earlier claims \cite{Devi:2013gya, Dhargyal:2016kwp, Dhargyal:2016jgo} that tensor interactions could explain the BaBar CP anomaly in $\tau \to K_S\pi \nu_\tau$ decays. This corresponds to the measurement of $A_{CP}=-3.6(2.3)(1.1)\times10^{-3}$ \cite{BABAR:2011aa}~\footnote{This observable is defined in section \ref{sec_ACP}.}, which disagrees remarkably with the SM prediction $A_{CP}=3.32(6)\times10^{-3}$, driven by neutral kaon mixing \cite{Bigi:2005ts, Calderon:2007rg}, probed with high accuracy in semileptonic kaon decays \cite{Tanabashi:2018oca}. In fact, the SM prediction is slightly modified by the experimental conditions corresponding to the reconstruction of the $K_S$ at the B-factory, yielding $A_{CP}=3.6(1)\times10^{-3}$ \cite{Grossman:2011zk}, which increases the discrepancy at the 2.8 $\sigma$ level. Although we will follow closely the results of Ref. \cite{Cirigliano-Crivellin}, as a novelty of our treatment, we will discuss the uncertainty induced on $A_{CP}$ by the error of the tensor form factor modulus.
 \item Three data points at the beginning of the $K_S\pi^-$ spectra measured by Belle \cite{Epifanov:2007rf} have been excluded from the reference fits or signalled as controversial in the dedicated analyses \cite{Moussallam:2007qc, Jamin:2008qg, Boito:2008fq, Boito:2010me, Antonelli-Cirigliano, Kimura:2014wsa, Bernard:2013jxa, Escribano:2014joa} and are at odds with the prediction \cite{Jamin:2006tk}. To our knowledge, only Ref.~\cite{Perez:2017mvc} was able to describe these data points due to the effect on the scalar form factor of the longitudinal correction to the $K^*(892)$ propagator induced by flavor symmetry breaking \footnote{As we will recall in section \ref{sec_Hadronization}, the scalar form factor contribution that we employ \cite{Jamin:2001zq} was obtained as a result of analyzing strangeness-changing meson-meson scattering \cite{Jamin:2000wn} within Chiral Perturbation Theory \cite{Gasser:1983yg, Gasser:1984gg} with resonances \cite{Ecker:1988te, Ecker:1989yg}, accounting for the leading flavor symmetry breaking.}. We will study if it is possible to explain these conflicting data points by the most general description of heavy NP contributions modifying the $\tau^-\to\bar{u}s\nu_\tau$ decays in the SM.
 \item Within an effective field theory analysis of possible non-standard charged current interactions, semileptonic tau decays \cite{Garces:2017jpz, Miranda-Roig, Cirigliano:2018dyk} have been proved competitive with the traditional semileptonic decays involving light quarks \cite{Cirigliano_Jenkins_Gonzalez, Bhattacharya_Cirigliano, Cirigliano:2012ab, Cirigliano:2013xha, Chang:2014iba, Courtoy:2015haa, Gonzalez-Alonso:2016etj, Gonzalez-Alonso:2016sip, Alioli:2017ces, Gonzalez-Alonso:2017iyc, Gonzalez-Alonso:2018omy}, like nuclear beta or leptonic and radiative pion decays. In this context, for the Cabibbo-suppressed sector, hyperon semileptonic decays \cite{Chang:2014iba, Gonzalez-Alonso:2016sip} cannot compete with (semi)leptonic Kaon decays \cite{Gonzalez-Alonso:2016etj} probing the scalar sector, given the (very accurately measured) dominant branching fractions of the latter and the suppressed ones (at most at the permille level) of the former. This intuitive reasoning suggests that strangeness-changing tau decays can also give non-trivial bounds on non-standard charged current interactions, although it is difficult that a competitive level with $K_{\ell(2,3)}$ decays \footnote{As we will discuss in section \ref{sec_Observables}, the quite limited momentum transfer of kaon and hyperon semileptonic decays makes them much cleaner theoretically than hadronic tau decays.}. The present work will make these statements precise.
\end{itemize}
We will answer the previous questions along the paper and summarize our conclusions in section \ref{sec_Concl}. The outline of the paper follows: in section \ref{sec_EFT} we present the effective Lagrangian describing the considered transitions, which is obtained from the general SMEFT Lagrangian after integrating non-dynamical degrees of freedom at $E\leq M_\tau$ and particularizing it for the Cabibbo-suppressed tau decays. In section \ref{sec_Amplitude} we introduce the lepton and hadron currents for the different SM and NP contributions and relate those of spins zero and one. For the meson part, we define the form factors that will be discussed at length in the section \ref{sec_Hadronization} next. These are built using chiral perturbation theory input into a dispersive framework and exploiting available measurements in order to reduce the associated uncertainties, which are important in our phenomenological analysis carried out in the remainder of this work. We focus on the tensor form factor, that turns out to be crucial to understand the strong $CP$ phase, through its interference with the dominant (accurately-known) vector form factor. Decay observables are discussed in section \ref{sec_Observables}, starting with the doubly differential Dalitz plot distributions (that we discuss for the first time, both in a pair of Mandelstam variables and also trading one of them by a measurable angle) in sections \ref{sec_Dalitz} and \ref{sec_Angular} and continuing with the single differential di-meson spectrum (in section \ref{sec_Rate}), which was measured by Belle \cite{Epifanov:2007rf}. Instead of presenting the angular distribution, we prefer to show the forward-backward asymmetry in section \ref{sec_AFB}. The limits on the NP  effective couplings are derived in section \ref{sec_Limits}, using: first only the branching ratio measurements of the $\tau^-\to(K\pi)^-\nu_\tau$ decays and then also the measured spectrum for the $K_S\pi^-$ decay channel. Our phenomenological analysis ends in section \ref{sec_ACP}, where we revisit \cite{Cirigliano-Crivellin} the $A_{CP}$ BaBar anomaly within the EFT framework.\\

\section{Effective theory analysis of $\tau^{-}\rightarrow\nu_{\tau}\bar{u}s$} \label{sec_EFT}
The lepton number conserving effective Lagrangian density constructed with dimension six operators and invariant under the local $SU(3)_C\otimes SU(2)_{L}\otimes U(1)_Y$ SM gauge group has the following form \cite{Buchmuller:1985jz, Grzadkowski:2010es},
\begin{equation}\label{Firsteq}
    \mathcal{L}^{(eff)}=\mathcal{L}_{SM}+\frac{1}{\Lambda^{2}}\sum_{i}\alpha_{i}O_{i}\longrightarrow \mathcal{L}_{SM}+\frac{1}{v^{2}}\sum_{i}\hat{\alpha}_{i}O_{i}\,,
\end{equation}
with $\hat{\alpha}_{i}=(v^{2}/\Lambda^{2})\alpha_{i}$ the dimensionless couplings encoding NP at a scale of some TeV.\\
In this framework we can explicitly construct the leading \footnote{We assume that higher-dimensional operators contributions, suppressed by powers of $M_\tau/M_W$, are negligible.} low-scale $\mathcal{O}$(1 GeV) effective Lagrangian (which has $SU(3)_C\otimes U(1)_{em}$ local gauge symmetry) for semi-leptonic transitions upon integrating the heavy degrees of freedom out \cite{Cirigliano_Jenkins_Gonzalez,Bhattacharya_Cirigliano},
\begin{equation}
    \begin{split}
    \mathcal{L}_{cc}=&\frac{-4G_{F}}{\sqrt{2}}V_{us}\Big[(1+[v_{L}]_{\ell\ell})\bar{\ell}_{L}\gamma_{\mu}\nu_{\ell L}\bar{u}_{L}\gamma^{\mu}s_{L}+[v_{R}]_{\ell\ell}\bar{\ell}_{L}\gamma_{\mu}\nu_{\ell L}\bar{u}_{R}\gamma^{\mu}s_{R}\\
    &+[s_{L}]_{\ell\ell}\bar{\ell}_{R}\nu_{\ell L}\bar{u}_{R}s_{L}+[s_{R}]_{\ell\ell}\bar{\ell}_{R}\nu_{\ell L}\bar{u}_{L}s_{R}\\
    &+[t_{L}]_{\ell\ell}\bar{\ell}_{R}\sigma_{\mu\nu}\nu_{\ell L}\bar{u}_{R}\sigma^{\mu\nu}s_{L}\Big] +\mathrm{h.c.}\,, \label{eqLcc}
        \end{split}
\end{equation}
where $G_{F}$ is the tree-level definition of the Fermi constant, $L(R)$ stand for left(right)-handed chiral projections and $\sigma^{\mu\nu}=i\left[\gamma^{\mu},\gamma^{\nu}\right]/2$. Note that if we set $v_{L}=v_{R}=s_{L}=s_{R}=t_{L}=0$, we recover the SM Lagrangian for the strangeness-changing semileptonic tau decays, with momentum transfer much smaller than the $M_W$ scale. Right-handed and wrong-flavor neutrino contributions were neglected in equation (\ref{eqLcc}) since they do not interfere with the SM amplitudes and do not contribute at leading order in the NP effective coefficients. The couplings $v_{L},v_{R},s_{L},s_{R},t_{L}$ are related to the $\alpha_i$ couplings of eq. (\ref{Firsteq}) in appendix A of Ref. \cite{Bhattacharya_Cirigliano} (assuming a weakly coupled scenario at the few-TeV scale). Besides Lorentz invariance, the only assumptions behind eq.~(\ref{eqLcc}) are the local gauge symmetries at low-energies ($U(1)_{em}$ and $SU(3)_C$ of quantum electrodynamics and chromodynamics, respectively) and the absence of light non-SM particles.\\
It is convenient to recast the spin-zero contributions in terms of currents with defined parity (scalar and pseudoscalar) in the following way
\begin{equation}\label{Efective_Lagrangian}
    \begin{split}
        \mathcal{L}_{cc}&=-\frac{G_{F}V_{us}}{\sqrt{2}}(1+\epsilon_{L}+\epsilon_{R})\Big[\bar{\tau}\gamma_{\mu}(1-\gamma_{5})\nu_{\ell}\cdot\bar{u}[\gamma^{\mu}-(1-2\hat{\epsilon}_{R})\gamma^{\mu}\gamma_{5}]s\\
        &+\bar{\tau}(1-\gamma_{5})\nu_{\ell}\cdot\bar{u}[\hat{\epsilon}_{s}-\hat{\epsilon}_{p}\gamma_{5}]s+2\hat{\epsilon}_{T}\bar{\tau}\sigma_{\mu\nu}(1-\gamma_{5})\nu_{\ell}\cdot\bar{u}\sigma^{\mu\nu}s\Big]+\mathrm{h.c.}\,,
    \end{split}
\end{equation}
where: $\epsilon_{L,R}=v_{L,R}$, $\epsilon_{s}=s_{L}+s_{R}$, $\epsilon_{p}=s_{L}-s_{R}$, and $\epsilon_{T}=t_{L}$. In eq.(\ref{Efective_Lagrangian}) we have particularized the Lagrangian for the tau lepton case $(\ell=\tau)$ \footnote{This implicit flavor dependence must be reminded for lepton universality tests, however.}, and we have also introduced the notation \\$\hat{\epsilon}_{i}=\epsilon_{i}/(1+\epsilon_{L}+\epsilon_{R})$ \cite{Garces:2017jpz} for $i=R,S,P,T$ \footnote{We note that this reshuffling is not convenient when comparing neutral and charged current processes and also when analyzing different semileptonic tau decays with an odd and an even number of pseudoscalar mesons, respectively \cite{Cirigliano:2018dyk}. Since $\epsilon_i=\hat{\epsilon}_i$ at linear order in these coefficients, we may use $\epsilon_i$ instead of $\hat{\epsilon}_i$ when comparing to works which use the former instead of the latter.}. In this way, our Lagrangian in eq.~(\ref{Efective_Lagrangian}) is equivalent to the one in eq. (9) of Ref.~\cite{Cirigliano-Crivellin} working at linear order in the epsilon Wilson coefficients. These are complex, although we will assume them real in first approximation in all subsequent analyses, but in the study of the CP asymmetry in section \ref{sec_ACP}, where a non-vanishing relative (weak) phase between the SM and tensor contributions is necessary to modify the SM result.\\

\section{Semileptonic $\tau$ decay amplitude} \label{sec_Amplitude}
In this section we calculate the decay amplitudes corresponding to the $\tau^{-}\rightarrow \bar{K}^{0} \pi^{-}\nu_{\tau}$ and the $\tau^{-}\rightarrow K^{-} \pi^{0}\nu_{\tau}$ decays. The first thing to note is that due to the parity of pseudoscalar mesons, only the vector, scalar and tensor currents give a non-zero contribution to the decay amplitude, as shown in the following equation  \footnote{Eq.(\ref{total_amplitude}) displays clearly that the renormalization scale dependence of the Wilson coefficients $\hat{\epsilon}_i$ needs to be cancelled by the one of the hadron matrix elements. As it is conventional, both are defined in the $\overline{MS}$ scheme at $\mu=2$ GeV.} ~\footnote{For convenience, the short-distance electroweak correction factor $S_{EW}$ \cite{Sirlin:1974ni, Sirlin:1977sv, Sirlin:1981ie, Marciano:1985pd, Marciano:1988vm, Marciano:1993sh, Braaten:1990ef, Erler:2002mv} is written as an overall constant, although it only affects the SM contribution. The error of this simplification is negligible working at leading order in the $\hat{\epsilon}_i$ coefficients \cite{Garces:2017jpz, Miranda-Roig}.}
\begin{equation}
\label{total_amplitude}
\begin{split}
\mathcal{M}=&\mathcal{M}_{V}+\mathcal{M}_{S}+\mathcal{M}_{T}\\
&=\frac{G_{F}V_{us}\sqrt{S_{EW}}}{\sqrt{2}}(1+\epsilon_{L}+\epsilon_{R})[L_{\mu}H^{\mu}+\hat{\epsilon}_{S}LH+2\hat{\epsilon_{T}}L_{\mu\nu}H^{\mu\nu}]\,,
\end{split}
\end{equation}
where the leptonic currents have the following structure ($p$ and $p'$ are the momenta of the tau lepton and its neutrino, respectively),\\
\begin{align}
\label{spinor_currents}
\begin{split}
L_{\mu}&=\bar{u}(p^{'})\gamma_{\mu}(1-\gamma_5)u(p)\,,\\
L&=\bar{u}(p^{'})(1+\gamma_5)u(p)\,,\\
L_{\mu\nu}&=\bar{u}(p^{'})\sigma_{\mu\nu}(1+\gamma_5)u(p)\,,\\
\end{split}
\end{align}
and the vector, scalar and tensor hadronic matrix elements for the case of the $\tau^{-}\rightarrow \bar{K}^{0} \pi^{-}\nu_{\tau}$ decay, are defined as follows
\begin{equation}
\label{vector_element}
H^{\mu}=\langle\pi^{-}\bar{K}^{0}|\bar{s}\gamma^{\mu}u|0\rangle=Q^{\mu}F_{+}(s)+\frac{\Delta_{ K\pi}}{s}q^{\mu}F_{0}(s)\,,
\end{equation}
\begin{equation}H=\langle\pi^{-}\bar{K}^{0}|\bar{s}u|0\rangle=F_{S}(s)\,,
\end{equation}
\begin{equation}
\label{tensor_matrix_element_1}
    H^{\mu\nu}=\langle\pi^{-}\bar{K}^{0}|\bar{s}\sigma^{\mu\nu}u|0\rangle=iF_{T}(s)(p^{\mu}_{K}p^{\nu}_{\pi}-p^{\mu}_{
    \pi}p^{\nu}_{K})\,,
\end{equation}
where $q^{\mu}=(p_{\pi}+p_{K})^{\mu}$, $Q^{\mu}=(p_{K}-p_{\pi})^{\mu}-\frac{\Delta_{K\pi}}{s}q^{\mu}$, $s=q^{2}$, and $\Delta_{ij}=m^{2}_{i}-m^{2}_{j}$. The hadron matrix elements $H,\,H^\mu$ and $H^{\mu\nu}$ were decomposed in terms of the allowed Lorentz structures, taking into account the discrete symmetries of the strong interactions, and a number of scalar functions of the invariant mass of the $K\pi$ system: the $F_S(s)$, $F_+(s)$, $F_0(s)$ and $F_T(s)$ form factors; which encode the details of the hadronization procedure.\\
The $\tau^{-}\rightarrow K^{-} \pi^{0}\nu_{\tau}$ decay is completely analogous. Neglecting (tiny) isospin corrections, the only  difference is given by the Clebsch-Gordan flavor symmetry factor of $\sqrt{2}$ between both decay channels, that is $\sqrt{2}F^{K^{-}\pi^{0}}_{0,+,T}(s)=F^{\bar{K}^{0}\pi^{-}}_{0,+,T}(s)$.\\
From equations (\ref{spinor_currents}) one can easily see that the vector and the scalar currents are related through the Dirac equation in the following way
\begin{equation}
    L=\frac{L_{\mu}q^{\mu}}{M_{\tau}}\,.
\end{equation}
Similarly, one can find a relation between the vector and the scalar hadronic matrix elements by taking the four-divergence of equation (\ref{vector_element}). This yields
\begin{equation}
    F_{S}(s)=\frac{\Delta_{K\pi }}{m_{s}-m_{u}}F_{0}(s)\,.
\end{equation}
Taking into account the previous two equations, we conclude that the scalar and vector contributions in eq.(\ref{total_amplitude}) can be treated jointly by doing the convenient replacement
\begin{equation}
    \frac{\Delta_{K\pi }}{s}\rightarrow  \frac{\Delta_{K\pi }}{s}\left[1+\frac{s\hat{\epsilon}_{s}}{M_{\tau}(m_{s}-m_{u})}\right]\,.
\end{equation}
Obtaining the three independent form factors ($F_0(s)$, $F_+(s)$ and $F_T(s)$) using as much experimental and theoretical knowledge as possible is the subject of the next section.\\

\section{Hadronization of the scalar, vector and tensor currents} \label{sec_Hadronization}
In this section we focus on the scalar, vector and tensor form factors. These are needed SM inputs for binding the non-standard interactions. Consequently, it is fundamental to obtain them reliably (including associated errors) so as to enable robust NP limits. This is conveniently done by using a dispersive framework (which warrants analyticity and elastic unitarity) that employs Chiral Perturbation Theory input (as given by low-energy QCD) and exploits available data to guide the dispersive integral and balance the precise theoretical knowledge at low-energies with the very accurate intermediate-energy measurements (especially near the resonance peaks). This approach is discussed minutely in refs. \cite{Jamin:2001zq, Boito:2008fq, Boito:2010me, Escribano:2014joa} for the scalar and vector $K\pi$ form factors. We will devote special attention to the corresponding dispersive tensor form factor  \cite{Cirigliano-Crivellin} following our earlier work \cite{Miranda-Roig} for the $\pi\pi$ case.\\

We will start with a brief reminder of the approach employed for the scalar form factor, $F_0(s)$. In a series of papers \cite{Jamin:2000wn, Jamin:2001zq, Jamin:2001zr, Jamin:2004re, Jamin:2006tj} Jamin, Oller and Pich studied strangeness-changing coupled-channels meson-meson scattering within Chiral Perturbation Theory with resonances and obtained very precise information on the corresponding scalar form factors, light quark masses and related chiral low-energy constants. We benefit from that analysis here \footnote{P. Roig thanks Matthias Jamin for providing him with these data.}. In particular, we employ the update presented in Ref. \cite{Jamin:2006tj} for the dispersive representation of the $K\pi$ channel, together with its corresponding uncertainties \footnote{For the analysis of the $K\pi$ spectra near threshold it is particularly important to employ a scalar form factor that is consistent with the information coming from S-wave $K\pi$ scattering (including the coupled channels $K\eta$ and $K\eta'$). The scalar form factor obtained in Ref. \cite{Jamin:2006tj} is included in the RChL version of TAUOLA \cite{Shekhovtsova:2012ra}, but not in other releases.}.\\

We turn next to the vector form factor. A dispersion relation for $F_{+}(s)$ was formulated in the references quoted previously. It is seen that a thrice-subtracted dispersion relation is optimal:
\begin{equation}\label{Vector-Dispersive}
    F_{+}(s)=\mathrm{exp}\left[\alpha_{1}s+\frac{\alpha_{2}}{2}s^{2}+\frac{s^3}{\pi}\int_{s_{\pi K}}^{\infty}ds^{'}\frac{\delta_{1}^{1/2}(s)}{(s^{'})^{3}(s^{'}-s-i\epsilon)} \right]\,,
\end{equation}
where $\alpha_{1}$, $\alpha_{2}$, and the one to set $F_{+}(0)=1$ are the three subtraction constants, and $s_{\pi K}=(m_{\bar{K}^{0}}+m_{\pi^{-}})^{2}$. Eq. (\ref{Vector-Dispersive}) shows that each additional subtraction in the dispersion relation gives rise to a further suppression factor $1/s^{'}$ in the integrand, enhancing the relative importance of the low-energy input.\\
In eq. (\ref{Vector-Dispersive}) Watson's final-state interactions theorem \cite{Watson:1952ji} was used. It states that below inelasticities the phase of the form factor equals the scattering phase of the K$\pi$ system ($\delta_{1}^{1/2}(s)$ in this case, as it has spin one and isospin one half). For this decay channel, departures are expected above $(m_K+m_\eta)^2\sim1.022$ GeV$^2$. These are accounted for in the analyses cited above and are included in our study. We consider as reference input the results obtained in section 3 of Ref. \cite{Boito:2010me} (without using constraints from Kaon decays \cite{Boito:2010me, Antonelli-Cirigliano} or information from $\tau^-\to K^-\eta\nu_\tau$ decays \cite{Escribano:2013bca, Escribano:2014joa}). The corresponding systematic and statistical errors of $F_+(s)$ that we use can be traced back to the results in Table 1 of Ref. \cite{Boito:2010me}: the (correlated) statistical errors of the fitted parameters characterizing $F_+(s)$ are those coming from the fit and given in this table and the systematic errors are estimated from the differences induced by changing $s_{cut}$ between the different columns of that table for a given $F_0(s)$. The phase $\delta_{1}^{1/2}(s)$ is confronted to data in Fig. 2 of Ref. \cite{Boito:2010me}. The fine agreement of $|F_+(s)|$ with the corresponding measurements can be appreciated in various plots of the papers quoted above.\\
A comment is in order with respect to the dependence on $s_{cut}$: Having a perfect knowledge of the form factor phase (including inelasticities) from threshold on, the dispersive integral should be performed with $s_{cut}\to\infty$. In this ideal case, a once-subtracted dispersion relation would suffice and $\alpha_{1,2,...}$ in eq.~(\ref{Vector-Dispersive}) could be obtained via sum rules and would perfectly match the coefficients of the polynomial terms of the low-energy expansion of $F_+(s)$. In practice, this is not the case, and one needs to increase the number of subtractions (three are optimal in this case, as we mentioned before). When this is done, a subtle cancellation (impossible to achieve exactly) is needed between this polynomial part and the one coming from the dispersive integral, to render a form factor that vanishes as $1/s$ asymptotically. That is why -even making the phase go to $\pi$ for $s\to\infty$- the asymptotic behaviour of the form factor does not comply exactly with the Brodsky-Lepage \cite{Lepage:1980fj} behaviour. Because of this shortcoming, performing the dispersive integral with $s_{cut}\to\infty$ is not unambiguously the best choice (this is discussed already in Ref. \cite{Jamin:2000wn}) as one would have expected. That is why we decided to follow Refs. \cite{Escribano:2013bca, Escribano:2014joa} and include the variation on $s_{cut}$ as the main source of systematic theory errors.\\

\begin{figure}
 \centering 
 \begin{subfigure}[b]{0.47\textwidth}
 \includegraphics[width=\textwidth]{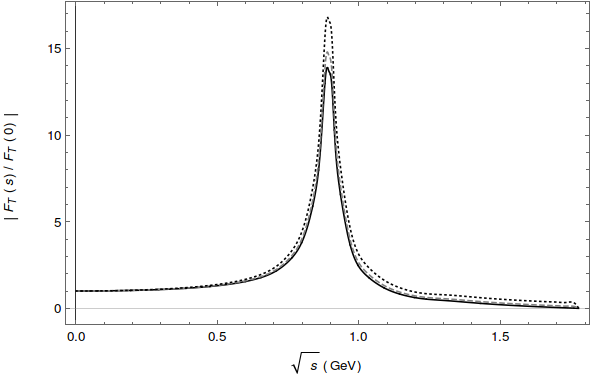}
  \end{subfigure}
 \begin{subfigure}[b]{0.47\textwidth}
  \includegraphics[width=\textwidth]{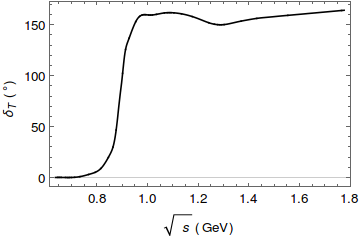}
   \end{subfigure}
   \caption{Modulus and phase, $|F_{T}(s)|$ (left) and $\delta_{T}(s)=\delta_+(s)$ (right), of the tensor form factor, $F_{T}(s)$. On the left plot, the dotted line corresponds to $s_{cut}=9$ GeV$^{2}$, the dashed one to $s_{cut}=4$ GeV$^{2}$, and the solid one to $s_{cut}=M^{2}_{\tau}$.}
   \label{figFT}
\end{figure}

To finish this section, we sketch the hadronization of the tensor current, which was presented in equation (\ref{tensor_matrix_element_1}), further details are given in Ref.~\cite{Miranda-Roig}. Its normalization at zero-momentum transfer is obtained using Chiral Perturbation Theory with tensor sources \cite{Cata-Mateu}. Its energy dependence is obtained solving numerically the dispersion relation, where the input phase corresponds to the one of the vector form factor in the elastic region \cite{Cirigliano-Crivellin}.\\
Proceedings analogously as in Ref.~\cite{Miranda-Roig}, we find that 
\begin{equation}
    i\left\langle \pi^{0}K^{-}\Big|\frac{\delta\mathcal{L}}{\delta\bar{t}_{\alpha\beta}}\Big|0\right\rangle=\frac{\Lambda_{2}}{\sqrt{2}F^2}
    \left(p^{\alpha}_{K}p^{\beta}_{0}-p^{\alpha}_{0}p^{\beta}_{K}\right)\,,
\end{equation} 
and verify the relative factor of $\frac{1}{\sqrt{2}}$ between the matrix elements for both decay channels.\\
The value of $\Lambda_{2}$ is not restricted by symmetry requirements and cannot be fixed from phenomenology. Fortunately, the lattice QCD evaluation of Ref.~\cite{Baum-Lubicz} found $f^{\bar{K}^0\pi^-}_{T}(0)=0.417(15)$. This, together with the fact that $F_{T}^{\bar{K}^0\pi^-}(0)=\frac{\Lambda_{2}}{F^{2}}$ implies that $\Lambda_{2}=(11.1\pm0.4)$ MeV, that we will use in our numerical analysis. This value is consistent within one sigma with the one employed in Ref. \cite{Miranda-Roig} for the $\pi\pi$ channel.\\
Unlike the vector and scalar form factor cases, there is no experimental data that can help us constructing $F_{T}(s)$ so that we must rely only on theory.\\
We calculate the energy-dependence of the tensor form factor $F_{T}(s)$ using again a phase dispersive representation as it is shown in refs. \cite{Cirigliano-Crivellin} and \cite{Miranda-Roig};
\begin{equation}\label{Tensor-Dispersive}
    \frac{F_{T}(s)}{F_{T}(0)}=\mathrm{exp}\left[\frac{s}{\pi}\int_{s_{\pi K}}^{\infty}ds^{'}\frac{\delta_{T}(s^{'})}{s^{'}(s^{'}-s-i\epsilon)}\right]\,,
\end{equation}
where $s_{\pi K}=(m_{\bar{K}^{0}}+m_{\pi^{-}})^{2}$. As in the scalar case we have included one subtraction. It is clear that, lacking precise low-energy information, we cannot increase the number of subtractions of $F_T(s)$. This, in turn, implies a sizable sensitivity to the upper limit of the integral that is used numerically ($s_{cut}$), which is illustrated in the left plot of our figure \ref{figFT}, where we consider the cases $s_{cut}=M_\tau^2, 4, 9$ GeV$^2$ \cite{Miranda-Roig}~\footnote{In principle, one could try to reduce this sensitivity following the strategies employed in Ref. \cite{SolisRoigpipiKK}, but the procedure will again be limited in this case by the absence of measurements sensitive to $F_T(s)$.}. We take the differences between these curves as an estimate of our systematic theoretical error on $F_T(s)/F_T(0)$. In the right panel of fig. \ref{figFT} we show the tensor form factor phase corresponding to $\delta_{T}(s)=\delta_+(s)$, with $\delta_+(s)$ from the fits in table 1 of Ref.~\cite{Boito:2010me}. In the inelastic region, our curve plotted for $\delta_{T}(s)$ lies within the error band shown in figure 2 of Ref.~\cite{Cirigliano-Crivellin} \footnote{Our phase is given in degrees while theirs is in radians.}.

The phases of $F_{T}(s)$ and $F_{+}(s)$ can be related as shown in Ref. \cite{Cirigliano-Crivellin}. We will not repeat their argument here, but only quote their main result: in the elastic region, $\delta_{T}(s)=\delta_{+}(s)=\delta^{1/2}_{1}(s)$. We will also estimate violations of this equation in the inelastic region (with their corresponding uncertainties) following again Ref. \cite{Cirigliano-Crivellin} (see fig. 2 in that reference).\\

\section{Decay observables} \label{sec_Observables}
In the rest frame of the $\tau$ lepton, the doubly differential decay width for the $\tau^-\to K_S \pi^-\nu_\tau$ process is
\begin{equation}\label{Differential_Width}
    \frac{d^{2}\Gamma}{dsdt}=\frac{1}{32(2\pi)^{3}M^{3}_{\tau}}\overline{|\mathcal{M}|^{2}}\,,
\end{equation}
where $s$ is the invariant mass of the $\pi^{-}K_S$ system taking values in the $(m_{K^0}+m_{\pi^-})^2\leq s\leq M^{2}_{\tau}$ interval, and
\begin{equation}
t^{\pm}(s)=\frac{1}{2s}\left[2s(M^{2}_{\tau}+m^{2}_{K^0}-s)-\left(M^{2}_{\tau}-s\right)\left(s+m^{2}_{\pi^-}-m^{2}_{K^0}\right)\pm\left(M^{2}_{\tau}-s\right)\sqrt{\lambda(s,m^{2}_{\pi^-},m^{2}_{K^0})}\right]\,, 
\end{equation}
with $\lambda(x,y,z)=x^{2}+y^{2}+z^{2}-2xy-2xz-2yz$ being the usual K\"allen function and $t=(P_\tau-p_\pi)^2$.

\subsection{Dalitz plots}\label{sec_Dalitz}
\begin{figure}
 \centering
 \begin{subfigure}[b]{0.47\textwidth}
 \includegraphics[width=\textwidth]{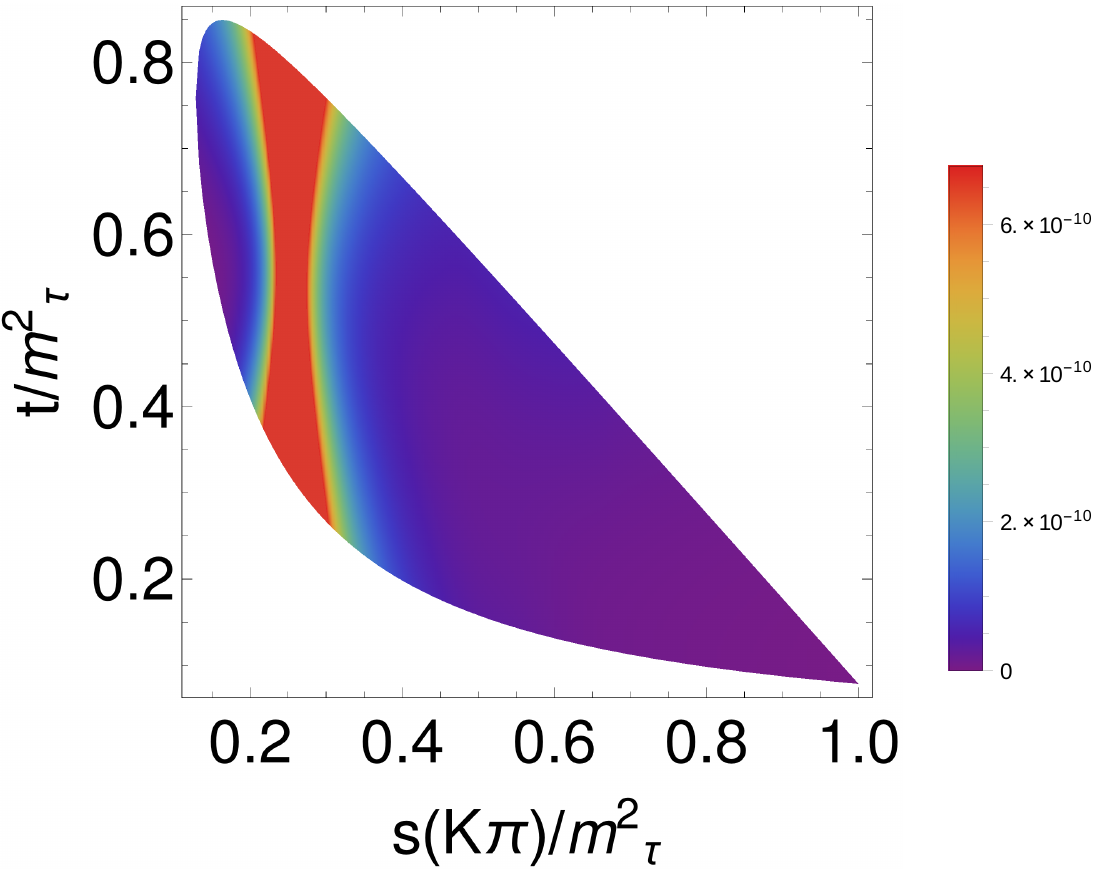}
  \end{subfigure}
 \begin{subfigure}[b]{0.47\textwidth}
  \includegraphics[width=\textwidth]{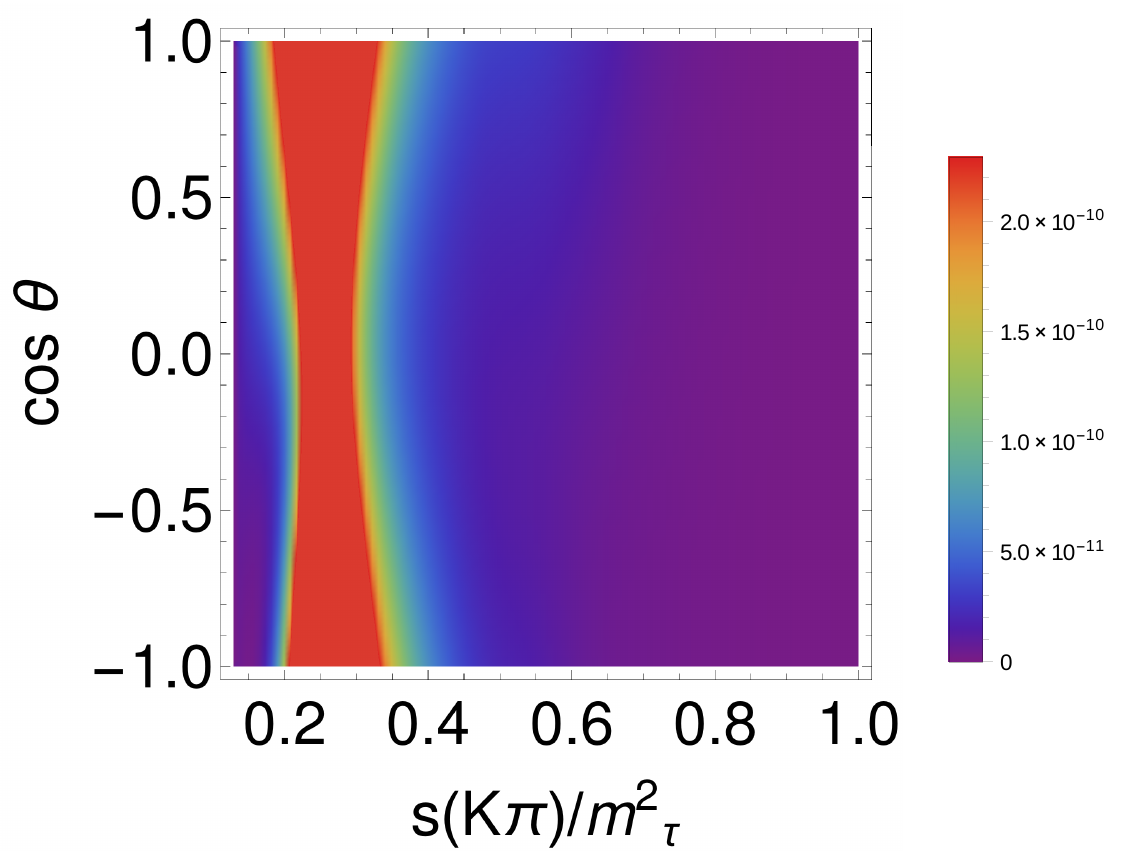}
   \end{subfigure}
   \caption{Dalitz plot distribution $\overline{|\mathcal{M}|^{2}}_{00}$ in the SM: Differential decay distribution for $\tau^{-}\rightarrow K_S \pi^{-}\nu_{\tau}$ in the $(s,t)$ variables (left). The right-hand figure shows the differential decay distribution in the $(s,\mathrm{cos}\theta)$ variables. The Mandelstam variables, $s$ and $t$, are normalized to $M^{2}_{\tau}$.}
   \label{Dalitz_SM}
\end{figure}

\begin{figure}[h]
 \centering
 \begin{subfigure}[b]{0.47\textwidth}
 \includegraphics[width=\textwidth]{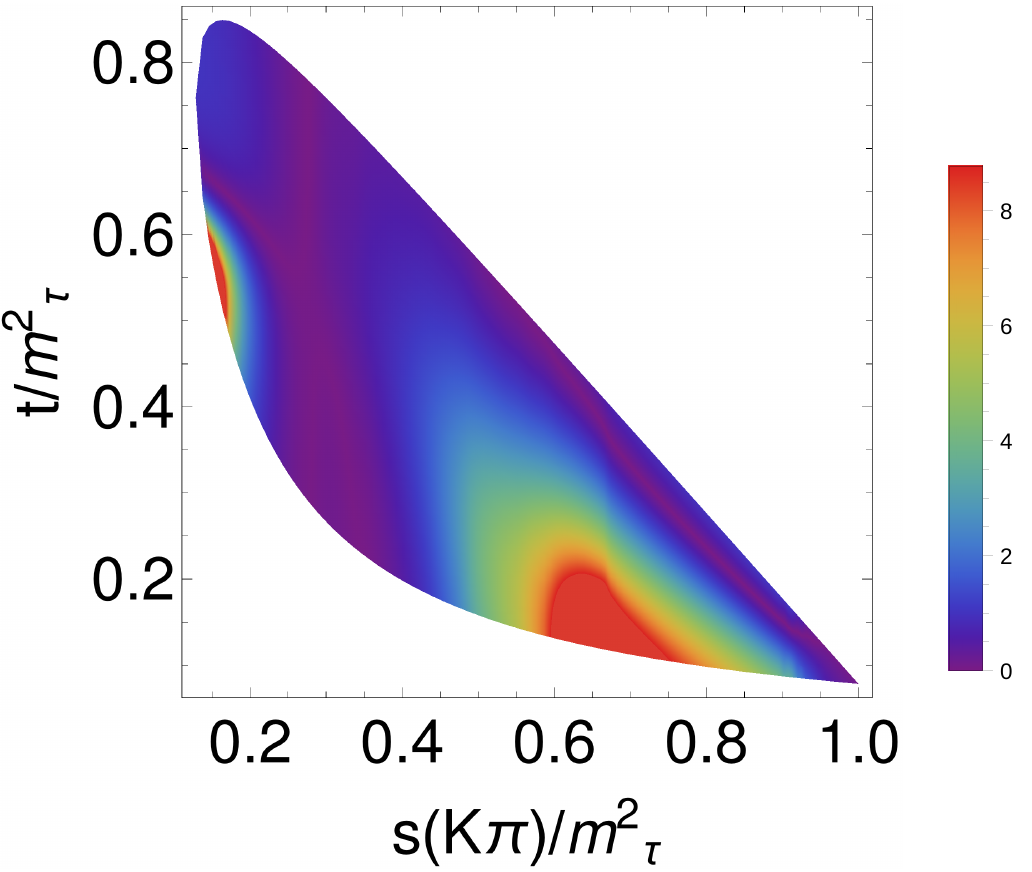}
  \end{subfigure}
 \begin{subfigure}[b]{0.47\textwidth}
  \includegraphics[width=\textwidth]{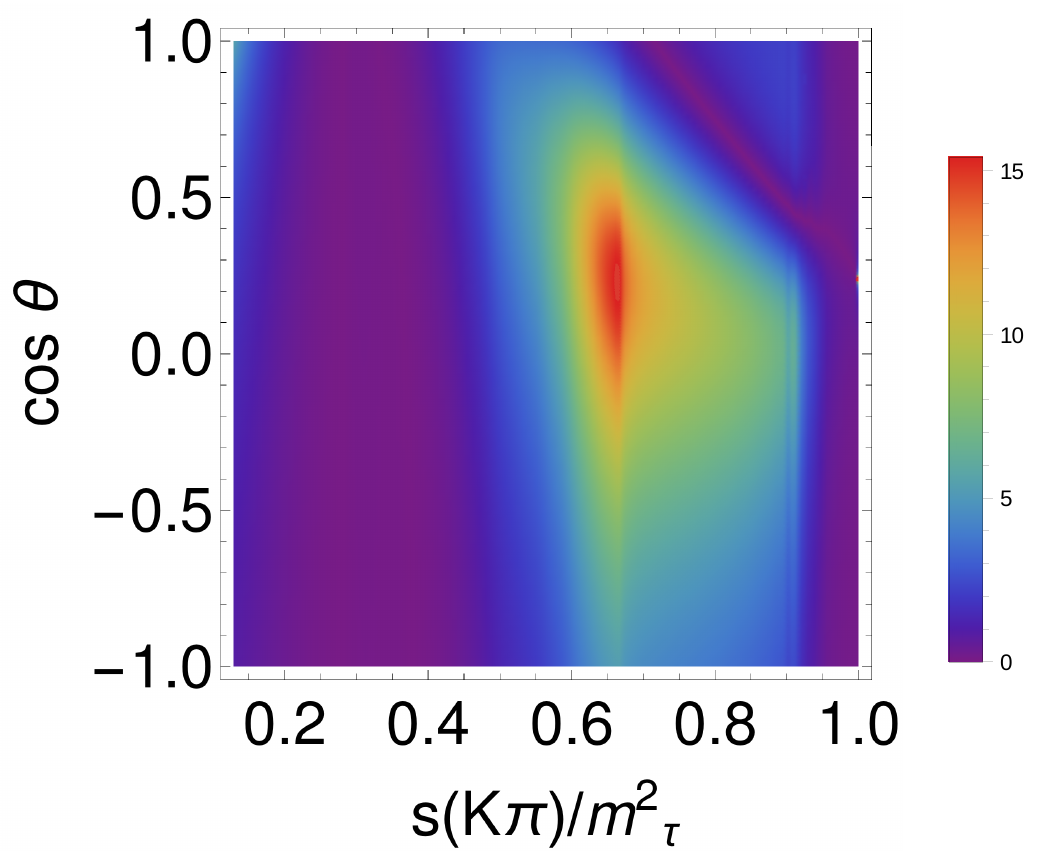}
   \end{subfigure}
   \caption{Dalitz plot distribution $\tilde{\Delta}(\hat{\epsilon}_{S},\hat{\epsilon}_{T})$, eq. (\ref{Delta_eps_tilde}), in the $\tau^{-}\rightarrow K_S \pi^{-}\nu_{\tau}$ decays: left-hand side corresponds to the differential decay distribution in the $(s,t)$ variables and the right-hand side to the one in the $(s,\mathrm{cos}\theta)$ variables, both with $(\hat{\epsilon}_{S}=-0.5,\hat{\epsilon}_{T}=0)$. The Mandelstam variables, $s$ and $t$, are normalized to $M^{2}_{\tau}$.}
   \label{Dalitz_eps_s}
\end{figure}

\begin{figure}[h]
 \centering
 \begin{subfigure}[b]{0.47\textwidth}
 \includegraphics[width=\textwidth]{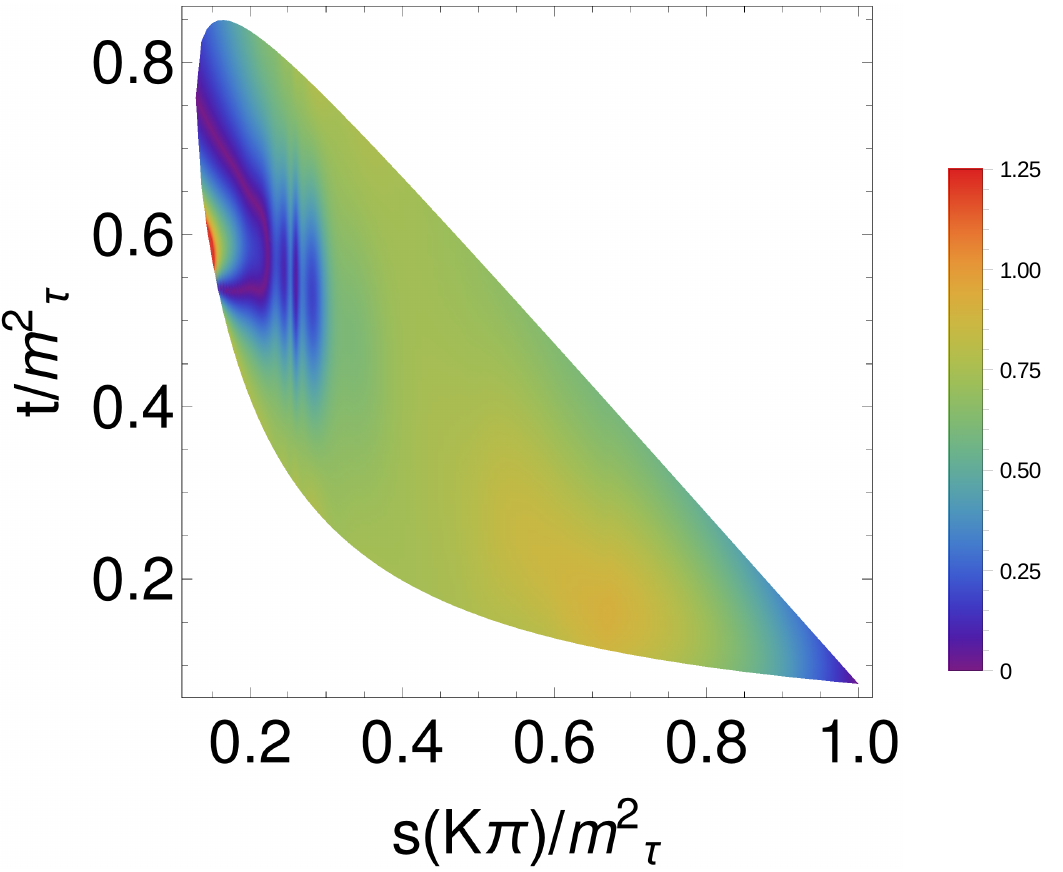}
  \end{subfigure}
 \begin{subfigure}[b]{0.47\textwidth}
  \includegraphics[width=\textwidth]{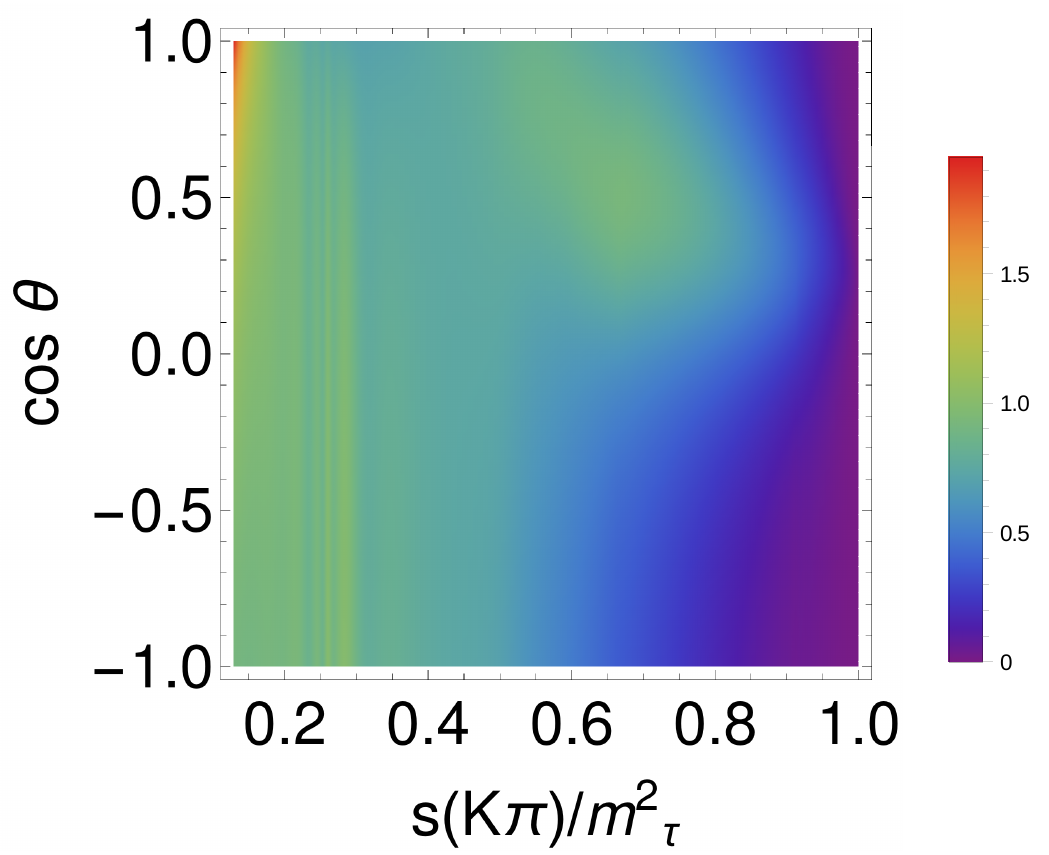}
   \end{subfigure}
   \caption{Dalitz plot distribution $\tilde{\Delta}(\hat{\epsilon}_{S},\hat{\epsilon}_{T})$, eq. (\ref{Delta_eps_tilde}), in the $\tau^{-}\rightarrow K_S\pi^{-}\nu_{\tau}$ decays: left-hand side corresponds to the differential decay distribution in the $(s,t)$ variables and the right-hand side to the one in the $(s,\mathrm{cos}\theta)$ variables, both with $(\hat{\epsilon}_{S}=0,\hat{\epsilon}_{T}=0.6)$. The Mandelstam variables, $s$ and $t$, are normalized to $M^{2}_{\tau}$.}
   \label{Dalitz_eps_t}
\end{figure}

Putting together the equations of section \ref{sec_Amplitude}, the amplitude (we will omit from now on the indices identifying the $K_S\pi^-$ charge channel) can be written as follows:
\begin{equation}\label{amplitude_parts}
    \mathcal{M}=\frac{G_{F}}{\sqrt{2}}V_{us}\sqrt{S_{EW}}(1+\epsilon_{L}+\epsilon_{R})(M_{0}+M_{+}+M_{T})\,,
\end{equation}
where it is straightforward to identify the precise form of the scalar ($M_{0}$), vector ($M_{+}$) and tensor ($M_{T}$) contributions from section \ref{sec_Amplitude}. All contributions to the squared matrix element (including interferences) are non-vanishing. Their expressions can be obtained from eqs. (5.4) in \cite{Miranda-Roig} taking into the corresponding $SU(3)$ Clebsch-Gordan coefficients and replacing $\Delta_{\pi^-\pi^0}\to-\Delta_{K\pi}$.\\
In the study of the Dalitz plots it is convenient to define the following observable introduced in Ref. \cite{Miranda-Roig}
\begin{equation}\label{Delta_eps_tilde}
\tilde{\Delta}(\hat{\epsilon}_{S},\hat{\epsilon}_{T})=\frac{\Big|\overline{|\mathcal{M}(\hat{\epsilon}_{S},\hat{\epsilon}_{T})|^{2}}-\overline{|\mathcal{M}(0,0)|^{2}}\Big|}{\overline{|\mathcal{M}(0,0)|^{2}}}\,,
\end{equation}
which is sensitive to the relative difference between the squared matrix element in presence/absence of NP  contributions (the SM case corresponds to $\mathcal{M}(0,0)$).\\
In the left panel of figure \ref{Dalitz_SM} we show the Dalitz plot for the SM case in the $(s,t)$ variables, and in the left part of figures \ref{Dalitz_eps_s} and \ref{Dalitz_eps_t} we show the corresponding plots for the values $(\hat{\epsilon}_{S}=-0.5,\hat{\epsilon}_{T}=0)$ and $(\hat{\epsilon}_{S}=0,\hat{\epsilon}_{T}=0.6)$, respectively. The election of these particular values of the $\hat{\epsilon}_{S,T}$ is discussed in section 5.5.\\

In the SM plots (figure \ref{Dalitz_SM}) it is clearly appreciated that the dynamics is dominated by the $K^*(892)$ vector  resonance but the effect of its excitation $K^*(1410)$ and of the dynamically generated $K^*_0(700)$ \cite{Buettiker:2003pp}, of  the $K^*_0(1430)$ and heavier states cannot be appreciated from the figure, although it is visible both in $F_+(s)$ and the decay  spectrum \cite{Boito:2008fq} and in $F_0(s)$ \cite{Jamin:2006tj}, respectively. The left panel of figures  \ref{Dalitz_eps_s} and \ref{Dalitz_eps_t} shows the relative modification of the squared matrix element for non-zero reasonable  values of $\hat{\epsilon}_S$ and $\hat{\epsilon}_T$ in the $(s,t)$ plane. Although large variations are seen in a couple of regions close to the border of the Dalitz plot in figure \ref{Dalitz_eps_s} (left), these correspond to zones with very suppressed probability, as can be seen in figure \ref{Dalitz_SM} (left). On the contrary, the regions with larger probability have a small relative change, according to figure \ref{Dalitz_eps_s} (left). In figure \ref{Dalitz_eps_t} (left) the region with the most noticeable change (though still smaller than those seen in figure \ref{Dalitz_eps_s}) is located very close to the $s$ minimum of the Dalitz plot, which has very small probability density in figure \ref{Dalitz_SM} (left). This region quite overlaps with one of the two mentioned for the fig. \ref{Dalitz_eps_s} left plot. Because of this feature, observing a deviation from the SM result in this region could be due to both tensor and non-standard scalar interactions. On the contrary, a deviation in the region of small $t$ values would be signalling spin-zero NP contribution. In any case, changes are very small in the region most densely populated by measured events in both left plots of figs. \ref{Dalitz_eps_s} and \ref{Dalitz_eps_t}. Due to this, we conclude that it will be extremely challenging to identify NP  contributions in the $(s,t)$ Dalitz plot even with the large data samples accumulated by the end of operation of Belle-II \cite{Kou:2018nap}.

\subsection{Angular distribution}\label{sec_Angular}
In this section we are going to study the angular dependence of the decay distribution. It is convenient to work in the rest frame of the hadronic system, in which we have $\vec{p}_{\pi}+\vec{p}_{K}=\vec{p}_\tau-\vec{p}_\nu=\vec{0}$, consequently the tau lepton and the pion energies are given by $E_{\tau}=(s+M_{\tau}^{2})/(2\sqrt{s})$ and $E_{\pi}=(s+m_{\pi}^{2}-m_{K}^{2})/(2\sqrt{s})$.\\
We will study the decay distribution in terms of the $(s,\mathrm{cos}\theta$) variables, where $\theta$ is the angle between the three-momenta of the pion and the three-momenta of the tau lepton, this angle is related to the invariant t variable by $t=M_{\tau}^{2}+m_{\pi}^{2}-2E_{\tau}E_{\pi}+2|\vec{p}_\pi||\vec{p}_{\tau}|\mathrm{cos}\theta$, where $|\vec{p}_{\pi}|=\sqrt{E_{\pi}^{2}-m_{\pi}^{2}}$ and $|\vec{p}_\tau|=\sqrt{E_{\tau}^{2}-M_{\tau}^{2}}$~\footnote{The tau lifetime and decay width ($\tau_\tau$ and $\Gamma_\tau$, respectively) are defined in the $\tau$ rest frame. Consequently, their values are boosted in the reference frame considered in this subsection.}.\\
The analytical results for the angular distribution can be obtained trivially from eqs. (5.6) in Ref. \cite{Miranda-Roig}, as mentioned above.\\
The Dalitz plots for the $(s,\mathrm{cos}\theta)$ variables are shown on the right panels of figures \ref{Dalitz_SM}, \ref{Dalitz_eps_s} and \ref{Dalitz_eps_t} (in these last two the observable $\tilde{\Delta}(\hat{\epsilon}_{S},\hat{\epsilon}_{T})$ is plotted). On figure \ref{Dalitz_SM} we plot the SM case, and in figures \ref{Dalitz_eps_s} and \ref{Dalitz_eps_t} we show Dalitz plots for the values $(\hat{\epsilon}_{S}=-0.5,\,\hat{\epsilon}_{T}=0)$ and $(\hat{\epsilon}_{S}=0,\,\hat{\epsilon}_{T}=0.6)$, respectively. The SM plot gives equivalent information in the $(s,\mathrm{cos}\theta)$ variables as the one seen in the $(s,t)$ variables (right versus left plot of figure \ref{Dalitz_SM}). Comparing both panels of figs. \ref{Dalitz_eps_s} one can see that one of the enhanced regions in the $(s,t)$ plot (the one at very low $s$ values) is washed away in the $(s,\mathrm{cos}\theta)$ diagram, while the other is slightly further enhanced in a limited region ($0\leq \mathrm{cos}\theta\leq 0.5$). The comparison of the left and right plots of figure \ref{Dalitz_eps_t} shows that the enhanced area for large $t$ values is a bit more prominent in the $(s,\mathrm{cos}\theta)$ distribution (for nearly maximal $\mathrm{cos}\theta$) although again it will be very hard to disentangle these possible deviations from the SM patterns in near future data.\\
Assuming approximate lepton universality, using the bounds from Ref. \cite{Gonzalez-Alonso:2016etj} (obtained analyzing Kaon (semi)leptonic decays) $\hat{\epsilon}_{S}\sim-8\times 10^{-4}$, $\hat{\epsilon}_{T}\sim6\times 10^{-3}$ (maximum allowed absolute values at one standard deviation) minimizes the deviations from the SM to unobservable level both in the $(s,t)$ and $(s,\mathrm{cos}\theta)$ Dalitz plots.\\

\subsection{Decay rate}\label{sec_Rate}
\begin{figure}[htp]
\centering
\includegraphics[scale=0.55]{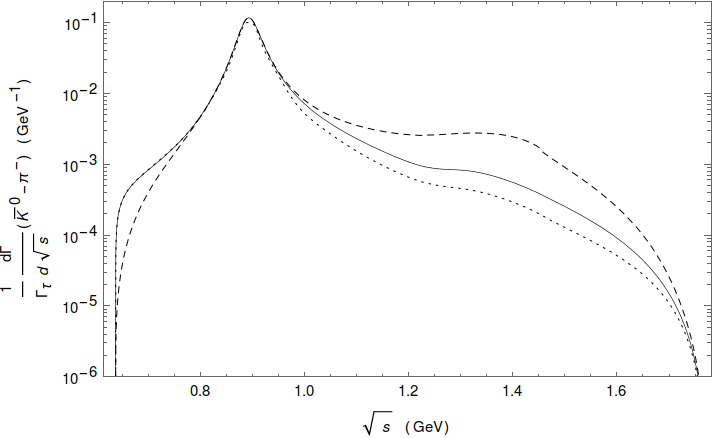}
\caption{The $\bar{K}^0\pi^{-}$ hadronic invariant mass distribution for the SM (solid line) and $\hat{\epsilon}_{S}=-0.5$, $\hat{\epsilon}_{T}=0$ (dashed line) and $\hat{\epsilon}_{S}=0$, $\hat{\epsilon}_{T}=0.6$ (dotted line). The decay distributions are normalized to the tau decay width.}
\label{Spectrum}
\end{figure}

Integrating eq. (\ref{Differential_Width}) upon the $t$ variable we obtain the invariant mass distribution, whose analytical form can be easily obtained from eqs. (5.7) and (5.8) in Ref. \cite{Miranda-Roig}.\\
It is important to mention that the only possible source of CP violation coming from the hadronic part is due to the Vector-Tensor interference, we will comment about this in section 6.\\
In figure \ref{Spectrum}, we plot the invariant mass distribution of the $K\pi$ system for $\tau^{-}\rightarrow K_S\pi^{-}\nu_{\tau}$ decays for the SM case and for $(\hat{\epsilon}_{S}=-0.5,\,\hat{\epsilon}_{T}=0)$ and $(\hat{\epsilon}_{S}=0,\,\hat{\epsilon}_{T}=0.6)$ which would be realistic values for these couplings, according to their impact on the decay width. Despite the logarithmic scale of the plot, the deviations from the SM curve shown in figure \ref{Spectrum} are too large when they are confronted with the Belle measurements of this spectrum, as we will see in the fits of section 5.5. This will allow us to set better bounds on $\hat{\epsilon}_{S,T}$ than those used in this subsection.
 
\subsection{Forward-backward asymmetry}\label{sec_AFB}
\begin{figure}[htp]
\centering
\includegraphics[scale=0.60]{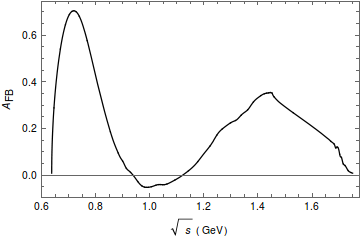}
\caption{Forward-backward asymmetry in $\tau^{-}\rightarrow K_S\pi^{-}\nu_{\tau}$ decays for the SM case.}
\label{FB_SM}
\end{figure}

\begin{figure}[htp]
\centering
\includegraphics[scale=0.60]{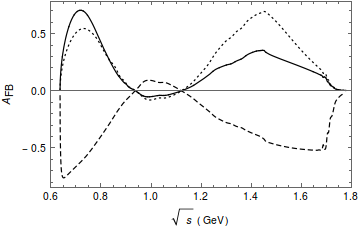}
\caption{Forward-backward asymmetry in $\tau^{-}\rightarrow K_S\pi^{-}\nu_{\tau}$ decays compared with the SM prediction (solid line). The dashed line corresponds to $\hat{\epsilon}_{S}=-0.5$, $\hat{\epsilon}_{T}=0$, and the dotted line corresponds to $\hat{\epsilon}_{S}=0$, $\hat{\epsilon}_{T}=0.6$.}
\label{FB_SM_es_et}
\end{figure}

\begin{figure}[htp]
\centering
\includegraphics[scale=0.66]{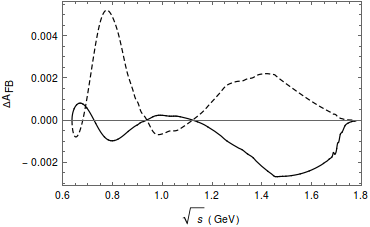}
\caption{Deviations from the SM forward-backward asymmetry, $\Delta\mathcal{A}_{K\pi}$, in $\tau^{-}\rightarrow K_S \pi^{-}\nu_{\tau}$ decays using the bounds from Ref. \cite{Gonzalez-Alonso:2016etj}. The solid line corresponds to $\hat{\epsilon}_{S}=-8\times10^{-4}$, $\hat{\epsilon}_{T}=0$ and the dashed line to $\hat{\epsilon}_{S}=0$, $\hat{\epsilon}_{T}=6\times10^{-3}$.}
\label{FB_SM_es_et_realistic_new}
\end{figure}

The forward-backward asymmetry is defined in analogy to the di-pion mode \cite{Miranda-Roig}
\begin{equation}\label{FB_Asym}
    \mathcal{A}_{K \pi}(s)=\frac{\int_{0}^{1}d\mathrm{cos}\theta\frac{d^{2}\Gamma}{dsd\mathrm{cos}\theta}-\int_{-1}^{0}dcos\theta\frac{d^{2}\Gamma}{dsd\mathrm{cos}\theta}}{\int_{0}^{1}d\mathrm{cos}\theta\frac{d^{2}\Gamma}{dsd\mathrm{cos}\theta}+\int_{-1}^{0}d\mathrm{cos}\theta\frac{d^{2}\Gamma}{dsd\mathrm{cos}\theta}}\,.
\end{equation}
Integrating the angular distribution upon the cos$\theta$ variable we obtain its analytical expression \footnote{In eq.~(\ref{FB_Asym_full}) we use $\mathcal{A}_{K\pi}$ to emphasize the decay channel under consideration and compare it next to our previous result for the $\pi\pi$ decay mode. Otherwise we will also be using the most common notation $A_{FB}$ for this observable.} 
\begin{equation}\label{FB_Asym_full}
\begin{split}
    \mathcal{A}_{K\pi}=&\frac{3\sqrt{\lambda(s,m^{2}_{\pi},m^{2}_{K})}}{2s^{2}[X_{VA}+\hat{\epsilon}_{S}X_{S}+\hat{\epsilon}_{T}X_{T}+\hat{\epsilon}^{2}_{S}X_{S^{2}}+\hat{\epsilon}^{2}_{T}X_{T^{2}}]}\left(1+\frac{s\hat{\epsilon}_{S}}{M_{\tau}(m_{s}-m_{u})}\right)\Delta_{\pi K}\\&\times\left[-Re[F_{0}(s)F^{*}_{+}(s)]+\frac{2s\hat{\epsilon}_{T}}{M_{\tau}}Re[F_{T}(s)F^{*}_{0}(s)]\right]\,.
    \end{split}
\end{equation}
Comparing to the $\pi\pi$ case, the main difference is that the very small parameter there $\Delta_{\pi\pi}$ is replaced here by $\Delta_{\pi K}$, which is not that small. The forward-backward asymmetry for the case in which $\epsilon_{R}=\epsilon_{L}=\hat{\epsilon}_{S}=\hat{\epsilon}_{T}=0$, corresponding to the SM, is plotted in figure \ref{FB_SM}. It should not be difficult to measure a non-zero forward-backward asymmetry around $\sqrt{s}\sim0.6$ GeV. Above the onset of inelasticities ($\sqrt{s}\gtrsim1.05$ GeV) the theory uncertainty starts to increase up to the kinematical upper limit of $\sqrt{s}$. It was already emphasized long ago that a measurement of the forward-backward asymmetry in this decay channel would be crucial in improving our knowledge of both vector and scalar form factors \cite{Beldjoudi:1994hi}~\footnote{We note that in this reference, and also later on in Refs. \cite{Gao:2012su, Kimura:2014wsa}, the angle $\theta$ used to compute $A_{FB}$ is defined between the three-momenta of the tau lepton and the $K_S$ in the di-meson rest frame. Taking into account the different sign conventions, it can be checked there is reasonable agreement with these works in the elastic region.}.\\
In figure \ref{FB_SM_es_et}, we plot $\mathcal{A}_{K\pi}$ for the values $(\hat{\epsilon}_{S}=-0.5,\,\hat{\epsilon}_{T}=0)$ and $(\hat{\epsilon}_{S}=0,\,\hat{\epsilon}_{T}=0.6)$, and we compare those plots with the SM case. For quite large $\hat{\epsilon}_{T}$ values some difference is appreciated for the tensor case; otherwise it may not be possible to disentangle it from the standard contribution. On the contrary, for non-standard scalar interaction, $A_{K\pi}$ flips sign with respect to the SM and it gets smaller in magnitude as $s$ increases. If it is possible to measure $A_{K\pi}$ in a low-energy bin, this would ease the identification of this type of NP in $A_{K\pi}$. When the more realistic limits obtained in Ref. \cite{Gonzalez-Alonso:2016etj} are considered (under the assumption of approximate lepton universality), it is impossible to identify any departures from the SM prediction in this observable. For this reason, we follow Ref.~\cite{Miranda-Roig} and use 
\begin{equation}
 \Delta \mathcal{A}_{K\pi}\,=\,\mathcal{A}_{K\pi}(s,\hat{\epsilon}_S,\hat{\epsilon}_T)-\mathcal{A}_{K\pi}(s,0,0)\,,
\end{equation}
instead. The corresponding (unmeasurably small) deviations from the SM result are plotted in figure \ref{FB_SM_es_et_realistic_new}.

\subsection{Limits on $\hat{\epsilon}_{S}$ and $\hat{\epsilon}_{T}$} \label{sec_Limits}
\begin{figure}[h!]
 \centering
 \begin{subfigure}[!htb]{0.49\textwidth}
 \includegraphics[width=\textwidth]{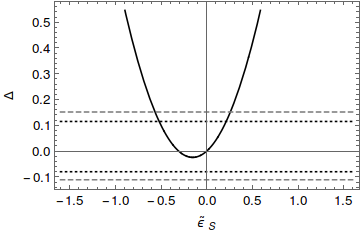}
  \end{subfigure}
 \begin{subfigure}[!htb]{0.49\textwidth}
  \includegraphics[width=\textwidth]{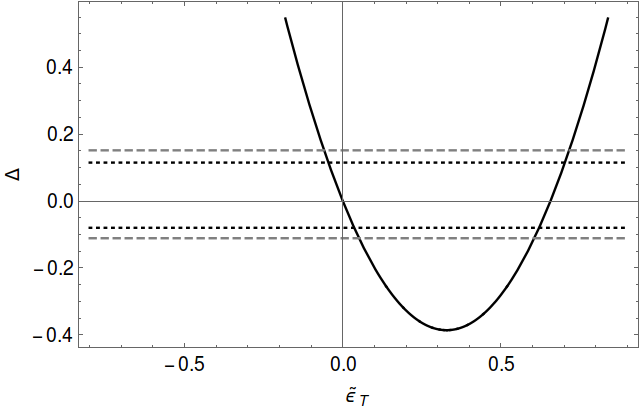}
   \end{subfigure}
   \caption{$\Delta$ as a function of $\hat{\epsilon}_{S}$ for $\hat{\epsilon}_{T}=0$ (left hand) and of $\hat{\epsilon}_{T}$ for $\hat{\epsilon}_{S}=0$ (right hand) for $\tau^{-}\rightarrow K_S \pi^{-}\nu_{\tau}$ decays. Horizontal lines represent the values of $\Delta$ according to the current measurement and theory errors (at three standard deviations) of the branching ratio (dashed line) and in the hypothetical case where the measured branching ratio at Belle-II has a three times reduced uncertainty (dotted line).}
   \label{Figure_Eps_S_Eps_T_1}
\end{figure}

One of the main purposes in the search for NP using the channel $\tau^{-}\rightarrow\bar{K}^{0}\pi^{-}\nu_{\tau}$ is to set bounds on the couplings $\hat{\epsilon}_{S}$ and $\hat{\epsilon}_{T}$, which are the effective couplings responsible of NP effects in this case. For this task we compare the decay width ($\Gamma$) for non-vanishing NP effective couplings with respect to the SM width ($\Gamma^{0}$) where NP is absent. We take the observable $\Delta$ defined in the following equation as the appropriate one to enhance the sensitivity to non-vanishing values of $\hat{\epsilon}_{S}$ and $\hat{\epsilon}_{T}$. 
\begin{equation}\label{NP_Shift}
    \Delta\equiv\frac{\Gamma-\Gamma^{0}}{\Gamma^{0}}=\alpha\hat{\epsilon}_{S}+\beta\hat{\epsilon}_{T}+\gamma\hat{\epsilon}^{2}_{S}+\delta\hat{\epsilon}^{2}_{T}\,,
\end{equation}
where we obtained the following results for the coefficients: $\alpha\in[0.30,0.34]$, $\beta\in[-2.92,-2.35]$, $\gamma\in[0.95,1.13]$ and $\delta\in[3.57,5.45]$. These values correspond to a slightly larger sensitivity to the possible  tensor contribution than to the effects of non-standard scalar interactions. We note that $|\beta|\sim\delta$, which corresponds to the fact that the interference between the spin-one SM amplitude and the tensor contribution is proportional to $Re[F_{T}(s)F^{*}_{+}(s)]$, which in the elastic region is $\sim |F_{T}(s)| |F_{+}(s)|$ and should be of similar magnitude than $|F_T(s)|^2$.\\
The values for $\hat{\epsilon}_{S}$ and $\hat{\epsilon}_{T}$ are calculated from eq. (\ref{NP_Shift}) in two different ways, as it is done in Refs. \cite{Garces:2017jpz, Miranda-Roig}. First we set one of the couplings to zero obtaining bounds for the other, these results are shown in figure \ref{Figure_Eps_S_Eps_T_1}.
\begin{figure}[h!]
\centering
\includegraphics[scale=0.48]{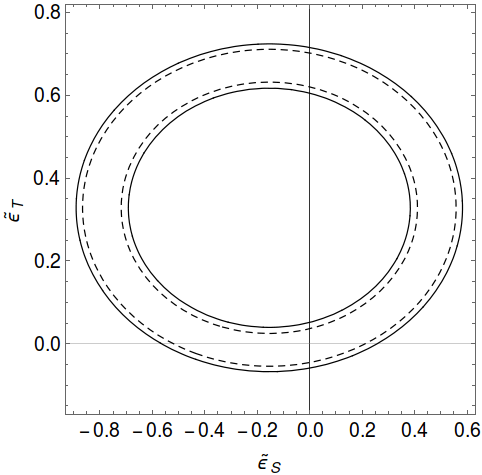}
\caption{Constraints on the scalar and tensor couplings obtained from $\Delta(\tau^{-}\rightarrow K_S\pi^{-}\nu_{\tau})$ using theory and the measured value reported in the PDG, with their corresponding uncertainties at three standard deviations (solid line). The dashed line ellipse corresponds to the case where the measurements error was reduced to a third of the current uncertainty.}
\label{elipse}
\end{figure}
Then, we also obtained constrains for $\hat{\epsilon}_{S}$ and $\hat{\epsilon}_{T}$ in the general case where both are non-vanishing. We show these results in figure \ref{elipse}, where the bounds on both couplings are limited by an ellipse in the $\hat{\epsilon}_{S}$-$\hat{\epsilon}_{T}$ plane.\\
The information for the couplings obtained here was used in the previous subsections, where we took the values $\hat{\epsilon}_{S}\sim-0.5$ and $\hat{\epsilon}_{T}\sim0.6$ as representative of realistic maximum absolute values of these coefficients when only the branching ratio (and not the decay spectrum) is considered.\\
Our results for the bounds in the $\hat{\epsilon}_{S}$ and $\hat{\epsilon}_{T}$ couplings are summarized in the following table.
\begin{table}[h!]
\label{table_1}
\begin{tabular}{|l|l|l|l|l|}
\hline
	$\Delta$ limits & $\hat{\epsilon}_{S}(\hat{\epsilon}_{T}=0)$ & $\hat{\epsilon}_{T}(\hat{\epsilon}_{S}=0)$ & $\hat{\epsilon}_{S}$ & $\hat{\epsilon}_{T}$\\
\hline
	Current bounds & $[-0.57,0.27]$ & $[-0.059,0.052]\cup[0.60,0.72]$  & $[-0.89,0.58]$ & $[-0.07,0.72]$\\
\hline

	Future bounds & $[-0.52,0.22]$ & $[-0.047,0.036]\cup[0.62,0.71]$ & $[-0.87,0.56]$ & $[-0.06,0.71]$\\
\hline

\end{tabular}
\caption{Constraints on the scalar and tensor couplings obtained through the limits on the current branching ratio at three standard deviations using the current theory and experimental errors and assuming the latter be reduced to a third ('Future bounds'). This last case should be taken only as illustrative of the improvement that can be achieved thanks to higher-statistics measurements, even in absence of any progress on the theory side. It is clear that the knowledge of $\hat{\epsilon}_{S,T}$ using $\tau^-\to K_S\pi^-\nu_\tau$ decays data is limited by theory uncertainties.}
\end{table}\\

The fact that the interval of allowed values for $\hat{\epsilon}_S$ is larger than for $\hat{\epsilon}_T$ (this is slightly more clear in the last two columns of the table, when both of them are floated simultaneously) agrees with our results for the $\beta$ and $\delta$ coefficients being larger (of similar magnitude, though, as the bounds on $\hat{\epsilon}_S$ and $\hat{\epsilon}_T$ are) than the $\alpha$ and $\gamma$ ones in eq. (\ref{NP_Shift}). According to the table, the allowed region is not symmetric and has a mild preference for $\hat{\epsilon}_S<0$ and $\hat{\epsilon}_T>0$.\\
Next we will consider fits to the branching ratio and decay spectrum~\footnote{P.~R. thanks Denis Epifanov for providing him with these data.} of the $\tau^-\to K_S\pi^-\nu_\tau$ decays as measured by Belle \cite{Epifanov:2007rf}. We will pay special attention to the possible explanation of the conflicting data points (bins $5$, $6$ and $7$) by the non-standard interactions. Therefore, we will consider fits with and without these data points. In all our fits, as explained e. g. in Ref.~\cite{Escribano:2014joa}, we will not consider the first data point (as it lies below the threshold for physical $K_S$ and $\pi^-$ masses) and will disregard the data from the last 10 bins, as suggested by the Belle collaboration.\\
The $\chi^2$ function minimized in our fits is
\begin{equation}
 \sum_i \left(\frac{{\mathcal{N}_i}^{exp}-\mathcal{N}_i^{th}}{\sigma_{\mathcal{N}_i}}\right)^2+\left(\frac{BR^{exp}-BR^{th}}{\sigma^{exp}_{BR}}\right)^2\,,
\end{equation}
where the sum over the $i$ bins may or may not include the $i=5,6,7$ bins. The theoretical expression for the differential decay rate (whose integration yields $BR^{th}\times\Gamma_\tau$) is related to the distribution of the measured number of events as indicated in eq. (3.1) of Ref. \cite{Escribano:2014joa} and in the subsequent explanation. We will consider the measurement of $BR^{exp}$ reported in the Belle paper \cite{Epifanov:2007rf} (and not the PDG \cite{Tanabashi:2018oca} or the HFLAV \cite{Amhis:2016xyh} values), as discussed in Ref.~\cite{Escribano:2014joa}. Along our fits we float the meson form factors within their estimated uncertainty bands and our quoted results take these errors into account. As in the EFT analysis of the $\tau^-\to\pi^-\pi^0\nu_\tau$ decays \cite{Miranda-Roig}, the data are not precise enough (and also are only for the di-meson spectrum and branching fraction, and not for the other observables) to allow a simultaneous fit of the parameters entering the dispersive representation of $F_+(s)$ and the NP coefficients $\hat{\epsilon}_{S,T}$ that improves the bounds obtained previously for the latter using only the branching ratio measurement.\\
First we have tried fits excluding the $i=5,6,7$ bins: initially varying $\hat{\epsilon}_S$ and $\hat{\epsilon}_T$ simultaneously \footnote{The correlation of $\hat{\epsilon}_S$ and $\hat{\epsilon}_T$ is always smaller than $0.4$ (in some fits it is even smaller than $0.1$) irrespective of including the $i=5,6,7$ bins in the fit or not. This is in agreeement with the very different energy-dependence of $F_0(s)$ and $F_T(s)$.}, with the results $\hat{\epsilon}_S=(1.3\pm0.9)\times10^{-2}$ and $\hat{\epsilon}_T=(0.7\pm1.0)\times10^{-2}$ and $\chi^2\sim[72,73]$, which improves very little with respect to the SM description, with a $\chi^2\sim[74,77]$ \footnote{The quoted intervals for the best $\chi^2$ include the systematic theoretical uncertainties as estimated varying $s_{cut}$ in Ref.~\cite{Boito:2008fq}. We are not following a more complete treatment of these uncertainties, as proposed in Ref.~\cite{SolisRoigpipiKK} for the $\pi^-\pi^0$ and $K^-K^0$ tau decays.}. Then we have set one in turn to values of the order of those found in Ref. \cite{Gonzalez-Alonso:2016etj} ($|\hat{\epsilon}_S|\leq1\times10^{-3}$ and $|\hat{\epsilon}_T|\leq6\times10^{-3}$, respectively) and fit the other one, with the results $\hat{\epsilon}_S=(1.4\pm1.6)\times10^{-2}$ (with fixed $\hat{\epsilon}_T$) and $\hat{\epsilon}_T=\left(0.9^{+1.0}_{-1.2}\right)\times10^{-2}$ (with fixed $\hat{\epsilon}_S$) with a $\chi^2$ of $[72,73]$ and $[73,74]$ respectively. These values are in accord with those of the joint fit, that we take as our reference results.\\
When we include the $i=5,6,7$ bins the previous results change in the following way: the reference SM result $\chi^2$ increases up to $\sim[91,95]$, and a simultaneous fit of both effective couplings yields $\hat{\epsilon}_S=(0.9\pm1.0)\times10^{-2}$ and $\hat{\epsilon}_T=(1.7\pm1.7)\times10^{-2}$, with a $\chi^2\sim[83,86]$. When we restrict $|\hat{\epsilon}_S|\leq1\times10^{-3}$ ($|\hat{\epsilon}_T|\leq6\times10^{-3}$), the best fit value of the floated coupling is $\hat{\epsilon}_T=\left(1.2^{+1.2}_{-1.0}\right)\times10^{-2}$ ($\hat{\epsilon}_S=(1.8\pm1.6)\times10^{-2}$) with a $\chi^2$ of $[86,89]$ ($[84,87]$). Again the single-parameter fits results agree with those obtained in the two-parameter fits, that we take as our central results.\\
All these fits favor slightly $\hat{\epsilon}_S>0$ and $\hat{\epsilon}_T>0$ (although negative $\hat{\epsilon}_T$ is possible within one sigma and $\hat{\epsilon}_S<0$ at less than $1.5$ standard deviations) \footnote{We recall that $\hat{\epsilon}_S<0$ and $\hat{\epsilon}_T>0$ were marginally preferred using only the branching ratio information.}. We summarize our main results in the following table.\\
\begin{table}[h!]
\label{table_2}
\begin{tabular}{|l|l|l|l|l|}
\hline
	Best fit values & $\hat{\epsilon}_{S}$ & $\hat{\epsilon}_{T}$ & $\chi^2$ & $\chi^2$ in the SM\\
\hline
	Excluding $i=5,6,7$ bins & $(1.3\pm0.9)\times10^{-2}$ & $(0.7\pm1.0)\times10^{-2}$  & $[72,73]$ & $[74,77]$\\
\hline
	Including $i=5,6,7$ bins & $(0.9\pm1.0)\times10^{-2}$ & $(1.7\pm1.7)\times10^{-2}$ & $[83,86]$ & $[91,95]$\\
\hline

\end{tabular}
\caption{Best fit values to the Belle spectrum and branching ratio of the $\tau^-\to K_S\pi^-\nu_\tau$ decays \cite{Epifanov:2007rf}. The cases where the $i=5,6,7$ bins are excluded/included are considered. We display the reference results obtained floating $\hat{\epsilon}_{S}$ and $\hat{\epsilon}_{T}$ simultaneously. In the last two columns the $\chi^2$ of these fits is compared to the SM result.}
\end{table}\\
In view of these results it is clear that the narrow peak structure constituted by the $i=5,6,7$ bins cannot be understood either in the SM (with a dispersive scalar form factor coming from the S-wave of a coupled channels analysis of meson-meson scattering \cite{Jamin:2000wn}) \cite{Jamin:2006tk, Moussallam:2007qc, Jamin:2008qg, Boito:2008fq, Boito:2010me, Antonelli-Cirigliano, Kimura:2014wsa, Bernard:2013jxa, Escribano:2014joa} or in the EFT analysis considered in this work. This conclusion agrees with the later preliminary data of BaBar \cite{Paramesvaran:2009ec} and a Belle posterior measurement \cite{Ryu:2013lca}, where such a bump near threshold is absent.\\

It would be desirable to compare these bounds to those obtained in (semi)leptonic Kaon \cite{Gonzalez-Alonso:2016etj} and hyperon \cite{Chang:2014iba} decays. In these cases, the momentum transfer corresponding to the hadronic transition is small enough that a dispersive treatment with chiral perturbation theory and lattice QCD input limits the systematic theory errors in a controlled way. Despite our careful treatment of the hadronic input used in this work, this cannot be the case for semileptonic tau decays because of the large range of the invariant mass of the hadron system. While our framework is able to provide an excellent control of the uncertainties in the elastic region, this is not the case out of it and we cannot discard that shortcomings of our approach (like for instance assuming that inelastic effects can be captured in the phaseshift that we employ, which is inherent to the elastic approximation) could mimic NP effects. This may be the case specially in the absence of alternative/complementary measurements to the Belle spectrum in $\tau^-\to K_S\pi^-\nu_\tau$ decays. With multi-dimensional information on these decays (through Dalitz plot analyses for instance) it should be possible -maybe requiring an improved theoretical description- to tell a NP scenario from a not sufficiently accurate description of the hadronic input. Taking this situation into account, our comparisons to (semi)leptonic Kaon and hyperon \cite{Chang:2014iba} decays should only be seen presently as tests of lepton universality.\\

The global fit of (semi)leptonic Kaon decays in Ref. \cite{Gonzalez-Alonso:2016etj} yielded the results $\hat{\epsilon}_S=(-3.9\pm4.9)\times10^{-4}$ and $\hat{\epsilon}_T=(0.5\pm5.2)\times10^{-3}$ for decays involving muons \footnote{We note that both signs for both coefficients are still possible at one standard deviation, so we find natural that our analysis cannot be conclusive on these signs either.}. If we assume (at least approximate) lepton universality we can compare the previous limits with the ones we have just obtained for the tau flavor which are $-0.89\leq\hat{\epsilon}_{S}\leq0.58$ and $-0.07\leq\hat{\epsilon}_{T}\leq0.72$ (when only the branching ratio is used) and $\hat{\epsilon}_S=(1.3\pm0.9)\times10^{-2}$ and $\hat{\epsilon}_T=(-0.7\pm1.0)\times10^{-2}$ (when the decay spectrum is also fitted). We verify that the $\tau^-\to (K\pi)^- \nu_\tau$ decays are not competitive with weak Kaon decays in binding $\hat{\epsilon}_S$ and that the corresponding lepton universality test is quite modest. In the case of $\hat{\epsilon}_T$ (where the limits from both types of decays differ only by a factor $3$) the  comparison agrees with lepton universality. Our bounds are at the same level of hyperon decays \cite{Chang:2014iba, Gonzalez-Alonso:2016sip}, where $|\epsilon_S|<4\times10^{-2}$ and $|\epsilon_T|<5\times10^{-2}$ are found at 90$\%$ confidence level. Therefore, while $\tau^-\to (K\pi)^- \nu_\tau$ decays cannot complement kaon decays concerning the bounds on $\hat{\epsilon}_{S}$ (neither hyperon decays do), the use of the three types of decays could allow to improve the knowledge of $\hat{\epsilon}_{T}$ (provided more complete and better measurements of $\tau^-\to (K\pi)^- \nu_\tau$ decays become available and these improvements are matched on the theory side).\\
Our results $\hat{\epsilon}_S=(1.6\pm0.9)\times10^{-2}$ and $\hat{\epsilon}_T=(0.9\pm1.0)\times10^{-2}$ translate into bounds on the corresponding NP scale $\Lambda\sim2-5$ TeV \footnote{Explicitly, $\Lambda\sim v (V_{us}\hat{\epsilon}_{S,T})^{-1/2}$, with $v=(\sqrt{2}G_F)^{-1/2}\sim246$ GeV.}, assuming effective couplings of natural value at $\mu=\Lambda$ and accounting for the running of these coefficients on the renormalization scale $\mu$ \cite{Broadhurst:1994se, Cirigliano:2012ab}. These results are, of course, quite restricted compared to the NP reach of (semi)leptonic Kaon decays, which can probe related scales as high as $\mathcal{O}(500)$ TeV \cite{Gonzalez-Alonso:2016etj} for non-standard scalar interactions.\\

\section{CP violation}\label{sec_ACP}
The observable $A_{CP}$, measured by BaBar \cite{BABAR:2011aa} at odds with the SM prediction (tiny corrections from direct CP violation are neglected along this section), is defined as

\begin{equation}
 A_{CP}=\frac{\Gamma(\tau^+\to\pi^+K_S\bar{\nu}_\tau)-\Gamma(\tau^-\to\pi^-K_S \nu_\tau)}{\Gamma(\tau^+\to\pi^+K_S\bar{\nu}_\tau)+\Gamma(\tau^-\to\pi^-K_S \nu_\tau)}\,.
\end{equation}
In the SM, $A_{CP}$ is saturated by the neutral kaon mixing contribution. Because of that, it is given by the analogous asymmetry measured in semileptonic kaon decays \cite{Cirigliano-Crivellin} ($\ell=e,\,\mu$)
\begin{equation}
 \frac{\Gamma(K_L\to\pi^-\ell^+\nu_\ell)-\Gamma(K_L\to\pi^+\ell^-\bar{\nu}_\ell)}{\Gamma(K_L\to\pi^-\ell^+\nu_\ell)+\Gamma(K_L\to\pi^+\ell^-\bar{\nu}_\ell)}=3.32(6)\times10^{-3}\,,
\end{equation}
up to small corrections given by the fact that the $K_S$ is reconstructed at the B-factories through its two-prong pion decay mode with a decay time of the order of the $K_S$ lifetime, which modify the previous value to $A_{CP}^{SM}=3.6(1)\times10^{-3}$ \cite{Grossman:2011zk}, that is 2.8 $\sigma$ away from the BaBar measurement, $A_{CP}=-3.6(2.3)(1.1)\times10^{-3}$.\\
In Ref.~\cite{Devi:2013gya} it is shown that -in the presence of beyond the SM (BSM) interactions- $A_{CP}$ is modified to
\begin{equation}
 A_{CP}=\frac{A_{CP}^{SM}+A_{CP}^{BSM}}{1+A_{CP}^{SM}\times A_{CP}^{BSM}}\,,
\end{equation}
where, in our case \cite{Cirigliano-Crivellin}~\footnote{We recall that $c_T$ in this reference equals $2\hat{\epsilon}_T$ in our notation.}
\begin{equation}\label{eqACPBSM}
 A_{CP}^{BSM}\,=\,\frac{2\mathrm{sin}\delta_T^W|\hat{\epsilon}_T|G_F^2|V_{us}|^2 S_{EW}}{256\pi^3M_\tau^2\Gamma(\tau\to K_S\pi\nu_\tau)}\int_{s_{\pi K}}^{M_\tau^2} \mathrm{d}s |f_+(s)||F_T(s)|\mathrm{sin}\left(\delta_+(s)-\delta_T(s)\right)\frac{\lambda^{3/2}(s,m_\pi^2,m_K^2)(M_\tau^2-s)^2}{s^2}\,,
\end{equation}
where $\delta_T^W$ stands for the relative weak phase between the SM V-A and the tensor contributions. In Ref.~\cite{Cirigliano-Crivellin}, using $SU(2)_L$ invariance of the weak interactions and the EFT machinery, poses stringent constraints on $\Im m[\hat{\epsilon}_T]$ by exploiting the measurements of $D-\bar{D}$ mixing and the upper limit on the electric dipole moment of the neutron. This results in the bound $2\Im m[\hat{\epsilon}_T]\lesssim10^{-5}$, that we will use. To see that $\delta_T^W$ is a small parameter, we remind the limits from the global EFT analysis of NP in Kaon (semi)leptonic decays \cite{Gonzalez-Alonso:2016etj}, according to which $|\epsilon_T|=(0.5\pm5.2)\times10^{-3}$. Considering this, $\mathrm{sin}\delta_T^W|\hat{\epsilon}_T|\sim\Im m[\hat{\epsilon}_T]$ and the numerical evaluation of eq.~(\ref{eqACPBSM}) is straightforward with the inputs at hand.\\
We have computed eq.~(\ref{eqACPBSM}) using $|F_T(s)|$ obtained with $s_{cut}=M_\tau^2,4,9$ GeV$^2$ (shown in the left panel of fig. 1) and with $\delta_T(s)$ varying (smoothly) within the band shown in fig. 2 of Ref. \cite{Cirigliano-Crivellin}, as we find perfectly reasonable the estimation of this uncertainty \footnote{Ref. \cite{Dighe:2019odu} disagrees with this and with the bound on $2\Im m[\hat{\epsilon}_T]$. See, however, Ref.  \cite{Roig:2019rwf}, where NP bounds obtained from $\tau^-\to K^-\nu_\tau$ decays are also discussed.}. The errors on $|F_+(s)|$ and $\delta_+(s)$ are negligible compared to the uncertainties on $F_T(s)$. Among these two types of uncertainties, the error on $\delta_T(s)$ dominates completely: changing $s_{cut}$ for a given $\delta_T(s)$ can modify $A_{CP}^{BSM}$ by a factor three, at most; while, with a fixed $s_{cut}$, $A_{CP}^{BSM}$ can be vanishing for $\delta_T(s)\to\delta_+(s)$ also in the inelastic region. In this way, we find 
\begin{equation}
A_{CP}^{BSM}\lesssim 8\cdot10^{-7}\,,
\end{equation}
which is slightly weaker bound than the one reported in Ref. \cite{Cirigliano-Crivellin}: $A_{CP}^{BSM}\lesssim 3\cdot10^{-7}$. This small difference comes mainly from our accounting for the variation in $s_{cut}$ and also for the slightly different phase $\delta_+(s)$ in both analyses. In any case, it is clear that heavy BSM interactions can only modify $A_{CP}$ at the $10^{-6}$ level at most \footnote{A similar upper limit was set in Ref. \cite{Delepine:2006fv} in the context of SUSY contributions to this observable. The recent paper \cite{Delepine:2018amd} finds that a charged Higgs contribution can enhance $CP$ violation in the $K^\pm\pi^0$ tau decays modes up to $A_{CP}\lesssim 2\times10^{-6}$, which is much larger than the corresponding (negligible) SM contribution \cite{Calderon:2007rg}.}, which is at least three orders of magnitude smaller than the theoretical uncertainty in its prediction (which is, in turn, some $25$ times smaller than the error of the BaBar measurement). Therefore, any conclusive anomaly in $A_{CP}$ must be explained outside the framework considered in this paper (and in Ref.~\cite{Cirigliano-Crivellin}); for instance, by BSM effects of very light particles.\\

\section{Summary and conclusions}\label{sec_Concl}

In this paper, besides giving an extended set of SM predictions for a bunch of observables (Dalitz plots, decay spectrum, forward-backward asymmetry) and comparisons with effects of heavy NP parametrized in terms of a couple of Wilson coefficients for the non-standard scalar and tensor interactions ($\hat{\epsilon}_{S,T}$) in $\tau^-\to (K\pi)^-\nu_\tau$ decays, we have focused on the three questions highlighted in section \ref{sec_Intro}, whose answers we summarize in the following:
\begin{itemize}
 \item In agreement with Ref. \cite{Cirigliano-Crivellin}, we confirm that it is not possible to understand within the low-energy limit of the SMEFT framework the BaBar measurement \cite{BABAR:2011aa} of the CP asymmetry, which disagrees at $2.8\sigma$ with the SM prediction \cite{Grossman:2011zk}. As a consequence of our dedicated treatment of the uncertainties on the tensor form factor, we find an slightly weaker bound than in Ref. \cite{Cirigliano-Crivellin}, $A_{CP}^{BSM}\lesssim 8\cdot10^{-7}$, which is anyway some three (five) orders of magnitude smaller than the theoretical uncertainty in its prediction (the error of the BaBar measurement). If the BaBar anomaly is confirmed, its explanation must be due to light NP. A determination of this quantity with Belle-I data, together with the future measurement at Belle-II \cite{Kou:2018nap}, will shed light on this puzzle.
 \item The bins number $5$, $6$ and $7$ of the Belle measurement \cite{Epifanov:2007rf} of the $K_S\pi^-$ mass spectrum in $\tau^-\to K_S\pi^-\nu_\tau$ decays could not find an explanation using a scalar form factor obtained from the corresponding partial-wave of a meson-meson scattering coupled channels analysis \cite{Jamin:2000wn, Jamin:2008qg} \footnote{The effect of the otherwise dominant vector form factor is kinematically suppressed in this region and can never give such a strong enhancement as observed in these data points.}. We have shown here, for the first time, that non-standard scalar or tensor interactions produced by heavy NP are not capable of explaining these data points either. Again a caveat remains with respect to light NP effects, which are beyond the scope of this paper.
 \item Current branching ratio and spectrum measurements of the $\tau^-\to K_S\pi^-\nu_\tau$ decays restrict the NP effective couplings, $\epsilon_S$ and $\epsilon_T$, as we have studied in this work for the first time. Our results are consistent with naive expectations: while the considered decays set bounds similar to those coming from hyperon semileptonic decays (which are at the level of a few TeV NP energy scale under reasonable assumptions), they are not competitive with (semi)leptonic Kaon decays, that could probe $\mathcal{O}(500)$ TeV NP generating non-standard scalar interactions. However, we put forward that tensor interactions are probed in $\tau^-\to (K\pi)^-\nu_\tau$ decays with similar NP energy reach than in (semi)leptonic Kaon and hyperon decays. Therefore, the corresponding comparisons for $\epsilon_T$ are meaningful tests of lepton universality and -under this assumption- tau decays can complement Kaon and hyperon physics in restricting tensor interactions.
\end{itemize}
We expect that the analysis carried out in this work can serve as a motivation for the experimental Tau Physics groups at Belle(-II) to undertake the very dedicated work leading to the measurement of the observables considered in this paper. According to our discussion (see also Ref.~\cite{Kou:2018nap}) the $CP$ asymmetry and the $K\pi$ spectrum are the most attractive measurements but it should never be forgotten that the power of correlations among observables can enhance the limits on (most interestingly light) NP coming from only these two measurements. As an example, a measurement of the corresponding forward-backward asymmetry can reduce the uncertainties of the SM input for both meson form factors, allowing thus to improve the limits in NP analyses.\\

\section*{Acknowledgements}P. R. thanks discussions with Sergi Gonz\`alez-Sol\'is concerning the estimation of the relevant form factors  uncertainties. This work has been partially funded by Conacyt: the support of project 250628  (Ciencia B\'asica), as well as the scholarship during J.~R. Ph. D. studies are acknowledged. The funding of Fondo SEP-Cinvestav 2018 is also recognized. G.~T. has been supported by a PIIF-IFUNAM project and CONACyT project 252167F.\\

\end{document}